\documentclass[sigconf]{acmart}
\AtBeginDocument{%
  \providecommand\BibTeX{{%
    \normalfont B\kern-0.5em{\scshape i\kern-0.25em b}\kern-0.8em\TeX}}}

\copyrightyear{2022}
\acmYear{2022}
\setcopyright{rightsretained}
\acmConference[FAccT '22]{2022 ACM Conference on Fairness, Accountability, and Transparency}{June 21--24, 2022}{Seoul, Republic of Korea}
\acmBooktitle{2022 ACM Conference on Fairness, Accountability, and Transparency (FAccT '22), June 21--24, 2022, Seoul, Republic of Korea}\acmDOI{10.1145/3531146.3533107}
\acmISBN{978-1-4503-9352-2/22/06}
\PassOptionsToPackage{numbers, sort, compress}{natbib}

\newcommand{\xhdr}[1]{\vspace{1mm} \noindent{\bf #1}.\:}
\newcommand{\qt}[1]{\textit{``#1''}}
\usepackage[suppress]{color-edits}
\usepackage{subfig}
\usepackage{float}
\usepackage{tabularx}
\usepackage{multirow}
\usepackage{wrapfig,stfloats}

\begin{document}

\title[Four Years of FAccT]{Four Years of FAccT: A Reflexive, Mixed-Methods Analysis of Research Contributions, Shortcomings, and Future Prospects}

\author{Benjamin Laufer}
\affiliation{%
   \institution{Cornell Tech}
   \country{United States}}
   \email{bdl56@cornell.edu}
   \orcid{0000-0002-3676-848X}
   
\author{Sameer Jain}\authornote{Equal contribution, ordered by seniority}
\affiliation{%
   \institution{Carnegie Mellon University}
   \city{Pittsburgh, PA}
   \country{United States}}
   \email{sameerj@andrew.cmu.edu}
   \orcid{0000-0003-1812-0425}

\author{A. Feder Cooper}
\authornotemark[1]
\affiliation{%
   \institution{Cornell University}
   \city{Ithaca, NY}
   \country{United States}}
   \email{afc78@cornell.edu}
   \orcid{0000-0002-4892-681X}
   
\author{Jon Kleinberg}
\affiliation{%
   \institution{Cornell University}
   \city{Ithaca, NY}
   \country{United States}}
   \email{kleinberg@cornell.edu}
   \orcid{0000-0002-1929-2512}
   
\author{Hoda Heidari}
\affiliation{%
   \institution{Carnegie Mellon University}
   \city{Pittsburgh, PA}
   \country{United States}}
   \email{hheidari@cmu.edu}
   \orcid{0000-0003-3710-4076}
   
\renewcommand{\shortauthors}{Laufer, et al.}
\renewcommand{\shorttitle}{Four Years of FAccT}
\begin{abstract}

Fairness, Accountability, and Transparency (FAccT) for socio-tech\-nical systems has been a thriving area of research in recent years. An ACM conference bearing the same name has been the central venue for scholars in this area to come together, provide peer feedback to one another, and publish their work. This \emph{reflexive} study aims to shed light on FAccT's activities to date and identify major gaps and opportunities for translating contributions into broader {positive} impact. To this end, we utilize a \emph{mixed-methods} research design. On the qualitative front, we develop a protocol for reviewing and coding prior FAccT papers, tracing their distribution of topics, methods, datasets, and disciplinary roots. We also design and administer a questionnaire to reflect the voices of FAccT community members and {affiliates} on a wide range of topics. On the quantitative front, we use the full text and citation network associated with prior FAccT publications to provide further evidence about topics and values represented in FAccT. We organize the findings from our analysis into four main dimensions: the \emph{themes} present in {FAccT scholarship}, the \emph{values} that underpin the work, the \emph{impact} of the contributions both within academic {circles} and beyond, and the practices and informal norms of the \emph{community} that has formed around FAccT. {Finally, our work identifies several} suggestions on directions for change, as voiced by community members.
\end{abstract}

\ccsdesc[500]{Social and professional topics~Computing / technology policy}
\ccsdesc[500]{Applied computing~Law, social and behavioral sciences}
\keywords{FAccT, mixed methods, reflexivity, community perspectives, topics, values, impact}

\maketitle
\section{Introduction} \label{sec:intro}
``Fairness, Accountability, and Transparency'' is a burgeoning area of research that examines the values embedded in socio-technical systems~\cite{facct2022webpage}. 
The area emerged amid concerns about the growing use of Artificial Intelligence (AI) and Machine Learning (ML) in socially consequential domains, and has evolved to include conferences, workshops, and books dedicated to this triptych of values.
With this growth has come the responsibility of the research community to contribute positively to broader social change~\citep{abebe2020roles}. 
The present work offers a reflexive view toward this scholarship, and attempts to provide a clearer picture of its emergence as a significant interdisciplinary field. In particular, we focus on the ACM FAccT conference, the flagship venue at the center of these research efforts.\footnote{FAccT has played a leading role in shaping the research agenda and it offers a concrete and meaningful way of specifying the boundaries of our inquiry. See \citep{laumann1989boundary} for a discussion of this type of \emph{boundary specification problem} in social-science
research.} We analyze FAccT's contributions and shortcomings, with an eye toward identifying fruitful directions for near-term improvements. 

Some scholars have argued that the community of researchers affiliated with FAccT has already made significant contributions, particularly by fostering numerous interdisciplinary interactions and raising awareness of the social, moral, and legal implications of technological work (see, e.g., ~\citep{chouldechova2018frontiers,kearns2019ethical,barocas-hardt-narayanan}). {At the same time}, there have been calls from within the {FAccT} community warning against several troubling trends---including a disproportionate focus on a handful of narrow topics (e.g., mathematical formulations of {outcome} fairness) at the expense of pressing challenges, such as AI governance~\citep{hutchinson201950, cooper2021eaamo, powles2018bias, abdurahman2019response, binns2018phil, cooper2021emergent, hellman2020measuring}. 
We aim to form a more nuanced and contextualized understanding of these views by consulting a variety of sources related to both data and methodology.

\xhdr{Reflexivity in scholarly field formation} This study takes a \emph{reflexive} stance towards scholarly inquiry \citep{bourdieu1992invitation}. \citet[p10]{bourdieu2000pascalian} describes reflexive scholarship as \qt{objectifying the subject of objectification,} by which he means \qt{deploying all the available instruments of objectification}---quantitative and qualitative empirical methods---to identify presuppositions, underlying values and assumptions.
Building on this notion, the present work puts forward a detailed analysis of FAccT research trends and themes to date with the goal of providing a foundation for broader community-wide discussions on the role and direction of the conference and scholarship. Our work takes a step toward igniting such constructive conversations by reflecting the collective voices of FAccT community members---including their views on the past and their recommendations for the future.
The authors of the present paper are themselves affiliated with FAccT, and in subsequent sections also reflect on the roles their own backgrounds play in the work.
As such, this analysis may serve as an example of reflexivity for young fields of scholarship \cite{bourdieu2004science}. This work is motivated in part by calls for data reflexivity from inside the FAccT community \cite{miceli2021documenting} as well as in the context of ML~\cite{elish2018situating}. 
Our work examines four critical pillars of FAccT: research \textbf{themes} embodied in its publications, \textbf{values} underlying the scholarship, \textbf{impact} (both intellectual and societal) that the work has garnered to date, and the practices and informal norms of the \textbf{community} of scholars who have come together through the venue. We ask:
\begin{itemize}
    \item \textbf{Themes (Section \ref{sec:themes}):} What are FAccT's main research topics and subtopics? 
    Which research approaches, methods and datasets are frequently utilized to examine these topics?  
    Which topics have been studied {more or less frequently}?
    Do researchers have sufficient access to `high quality' datasets? 
    \item \textbf{Values (Section \ref{sec:values}):} What are the {values} underpinning FAccT publications? How has the community interpreted fairness, accountability, and transparency?   
    Are there social values, moral foundations, ethical principles, and political ideologies that FAccT can address more extensively in the future?
    \item \textbf{Impact (Section \ref{sec:impact}):} To what extent have FAccT publications generated real-world and scholarly impacts?
        Has FAccT's intellectual influence been {local} (e.g., within intellectual echo chambers) or broad (e.g., bridging disciplinary gaps)? 
    \item \textbf{Community (Section \ref{sec:community}):} How do affiliates assess FAccT as a scholarly community? {How do they perceive the practices, informal norms, and academic culture of the FAccT community?}
\end{itemize}
We utilize a mix of qualitative and quantitative methods to elicit answers to the above questions~(Section~\ref{sec:methodology}). From these analyses we obtain several insights:
Our \textbf{thematic} exploration of FAccT publications shows that there has been an out-sized focus on ML as a computing subfield, and in quantitative work on fairness, displacing discussions about broader AI policy and governance (Section~\ref{sec:themes}).
We identify a list of \textbf{values} beyond fairness, accountability and transparency that are currently less well-represented at FAccT~(Section~\ref{sec:values}). While our analysis does not attempt to describe the set of end-to-end deployments in practice, we report community views on deployment and broader \textbf{impact}, which are generally believed to be less than expected. We do find, however, that FAccT has played a positive role in exposing its community to insights and ideas from other disciplines~(Section~\ref{sec:impact}).
Lastly, the \textbf{community}'s self assessment additionally reveals concerns around the practices and informal norms of the conference, including peer review {practices, lack of inclusivity, and out-sized industry connections} (Section~\ref{sec:community}).

\xhdr{Recommendations} To overcome some of the identified challenges, FAccT affiliates suggested several steps, including: (1) The conference organization should (perhaps ironically) be more transparent, especially concerning relationships with industry and aspects of the peer review process.
(2) The community needs to foster a more inclusive environment, including (but not limited to) appreciating various modes of inquiry and forms of contributions.
(3) The researchers must work closely with stakeholders and practitioners to have positive, real-world impact.
(4) The scholarship must critically evaluate the assumptions it takes for granted, for example, by developing a more standard set of terms and norms that do the necessary translational work for the conference and make its communications more effective.

\section{Our Mixed-Methods Design}\label{sec:methodology}

\begin{table}
\caption{Overview of our mixed-methods and datasets} 
\vspace{-0.075in}
\label{table:methods}
\small
\begin{tabular}{l l l l}
\toprule \textbf{Approach}        & \textbf{Method} & \textbf{Dataset}  & \textbf{Paper sections} \\ 
\midrule
\shortstack[l]{Qualitative\; \\ 
                           \; \\ 
                           \;  \\
                           \;} &
\shortstack[l]{Manual\; \\ 
               coding\; \\ 
               \; } & 
\shortstack[l]{FAccT pubs. corpus,\; \\ 
               survey responses \; \\ 
               \;} &  
\shortstack[l]{Themes (\ref{sec:themes}),\; \\ 
               values (\ref{sec:values}),\; \\ 
               community (\ref{sec:community})\;} \\ 
\midrule

\shortstack[l]{\\
Qualitative\; \\ 
                           \; \\ 
                           \; \\
                           \;} &
\shortstack[l]{\\
Survey\; \\ 
               \; \\ 
               \; \\
               \;} & 
\shortstack[l]{Responses to open-\; \\ 
               ended questions\; \\ 
                           \;} &  
\shortstack[l]{Values (\ref{sec:values}),\; \\ 
               impact (\ref{sec:impact}),\; \\ 
               community (\ref{sec:community})\;} \\ 
\midrule

\shortstack[l]{Quantitative\; \\ 
                           \;} &
\shortstack[l]{Survey\; \\ 
               \; } & 
\shortstack[l]{Responses to multi-\; \\ 
               choice questions\;} &
\shortstack[l]{Impact (\ref{sec:impact})\; \\ 
               \;} \\ 
\midrule

\shortstack[l]{Quantitative\; \\ 
                           \;} &
\shortstack[l]{Network\; \\ 
               analysis\; } & 
\shortstack[l]{Citation network\; \\ 
               of FaccT articles\;} &
\shortstack[l]{Themes (\ref{sec:themes})\; \\ 
               \;} \\ 
\midrule

\shortstack[l]{Quantitative\; \\ 
                           \;} &
\shortstack[l]{Topic\; \\ 
               modeling\; } & 
\shortstack[l]{FAcct pubs. corpus\; \\ 
              \;} &
\shortstack[l]{Themes (\ref{sec:themes})\; \\ 
               \;} \\ 
\midrule               
\shortstack[l]{Quantitative\; \\ 
                           \; \\ \;} &
\shortstack[l]{Moral \;\\
               founda-\; \\ 
               tions dict.\; } & 
\shortstack[l]{FAcct pubs. corpus\; \\ 
              \; \\
              \;} &
\shortstack[l]{Values (\ref{sec:values})\; \\ 
               \; \\ \;} \\ 
\bottomrule
\end{tabular}
\vspace{-.15in}
\end{table}
This study follows a mixed-method design, summarized in Table~\ref{table:methods}: We collect both qualitative and quantitative data related to our research questions, and we utilize qualitative and quantitative methods to analyze the resulting data. Our motivation for using mixed methods is two-fold: 1) we want to present a comprehensive picture of FAccT; drawing on both qualitative and quantitative methods helps achieve this, as each helps address limitations of the other. 2) our hope is for this study to offer actionable guidelines to FAccT; given the multidisciplinary background of FAccT scholars, it is appropriate to use methods that not only capture, but also reflect, this diversity. We begin by reflecting on our roles as researchers. We present overviews of our methods (Table~\ref{table:methods}) in Section~\ref{sec:methods_overview} and a description of our survey design in Section~\ref{sec:survey}. Further details on our use of particular methods can be found in relevant sections.

\xhdr{Reflecting on our roles as researchers}
In qualitative research, the researcher is the key instrument for gathering and making sense of data. So their background and motives (cultural, disciplinary, personal, ethical, strategic, or otherwise) play an essential role in shaping the direction and outcome of their research. As such, it is paramount that they reflect explicitly on the potential influence of their background, biases, and values on the research process~\citep{creswell2017research}. {Following this tradition, we next provide information about our backgrounds and reflect on our roles as researchers.} 

Our team consists of five researchers, two in senior and three in junior roles. Collectively, our team represents a variety of gender identities, ethnicities, and cultural, socio-economic, and national backgrounds. However, our team does not represent a broad range of political views. We all identify with liberal, progressive and/or left-wing (as opposed to conservative) ideals and values. We are currently affiliated with academic institutions in the U.S., and some of us have worked in the industry in the past. Three of us have had sustained interest and involvement in the FAccT scholarship. We generally hold favorable views toward the conference and the affiliated research community. However, our past experiences and perceptions regarding the lack of diversity in topics, backgrounds, and politics represented at FAccT motivated us to undertake the current study. We believe our close affiliation with FAccT elevates our understanding of the research landscape, and our sensitivity and care toward the challenges faced by FAccT affiliates and stakeholders. The background and experiences described above have undoubtedly shaped our choice of research questions and our interpretations of the data. For example, as mentioned earlier, we initiated the study with the conviction that the community would benefit from a more diverse representation of issues, politics, and research paradigms. Our position and background have also provided us with various forms of access---to citation data from AMiner and Semantic Scholar, to conference proceedings, and above all, to FAccT community members who agreed to participate in our questionnaire and share their reflections and recommendations.

\subsection{Overview of Methods}\label{sec:methods_overview}

\noindent\textbf{Coding} is a method of organizing qualitative data ``comprised of processes that enable collected data to be assembled, categorized, and thematically sorted, providing an organized platform for the construction of meaning''~\cite[p45]{williams2019art}. A code is often a ``short phrase that symbolically assigns a summative, salient, essence-capturing, and/or evocative attribute for a portion of [...] data''~\citep[p3]{saldana2021coding}. We use manual coding in our thematic analysis of FAccT publications, and in interpreting FAccT affiliates' responses to open-ended questions about shortcomings and recommendations. {Further details about our coding protocols will be presented in the appropriate sections.}

\noindent\textbf{Topic modeling}, particularly the Bayesian unsupervised learning technique of Latent Dirichlet Allocation (LDA)~\cite{blei2003lda}, is a popular and well-documented tool in natural-language processing (NLP) for eliciting thematic information from a text corpus. LDA operates on a bag-of-words representation of text documents and enables us to model each FAccT paper as a distribution over topics, where each topic is a distribution over vocabulary words. A topic can be understood as a set of frequently-present, co-located words, based on which we can assign a semantically meaningful overarching topic label~\cite{chang2009tealeaves}. From our model, we determine which papers belong to which topics; this enables us to develop a quantitative understanding of prominent research themes in FAccT---year-over-year changes and across all four years.

\noindent\textbf{Community detection} in citation networks is an unsupervised quantitative technique that enables us to split a citation graph into subgraphs (called \textit{communities}), such that the nodes within a subgraph have denser connectivity, share properties, or play similar roles within the graph. We use the Louvain community detection algorithm~\cite{blondel2008fast} to elicit communities within the extended FAccT citation network---comprised of FAccT publications and their immediate citation connections. By definition, the papers within a community exhibit a higher concentration of citation relationships than those across communities~\cite{fortunato2010community}, so we expect them to roughly map to sub-areas of research. We analyze titles of the papers that appear in each community to assign an overarching thematic label to it. 
We contrast the outcome of the above approaches to the thematic exploration of FAccT in Section~\ref{sec:themes}.

\noindent\textbf{Surveys} directly solicit data from the population of interest---in our case, FAccT affiliates. By asking both open-ended and {multiple-choice, Likert-scale rating questions \cite{likert1932technique}}, our web questionnaire {aims to} gather candid, less-biased views from participants \cite{reja2003open}. Open-ended questions, a mode of structured interview, invite qualitative research analysis stemming from \textit{phenomenology}. This style of analysis tries to identify and clarify phenomena as they are experienced by individuals, rather than from an abstract or objective perspective~\cite{groenewald2004phenomenological}. Closed-ended questions, which can be categorized under \textit{quantitative survey design}, are suitable to answer descriptive research questions about the relationship between variables of interest---in our case, FAccT's topics and impact. The next Section provides details about our survey design.

\subsection{A Mixed-methods Survey of FAccT Community Members}\label{sec:survey}
{An} essential component of our analysis is a web-based {survey} designed to solicit FAccT affiliates' responses to questions in three broad categories: 
1) views on FAccT scholarship and recommendations for future improvements;
2) intellectual merit and broader impact of several FAccT research topics;
3) broader impact of FAccT scholarship in several application domains.
The questionnaire ended with an optional set of questions about the participants' background and affiliation with FAccT. 
The {qualitative component of our survey} contained four main open-ended questions: 
\textit{
\begin{itemize}
    \item ``Are there any moral or social values (sufficiently distinct from Fairness, Accountability, and Transparency) that you believe FAccT scholarship should address in near future?''
    \item ``What do you consider to be the most important criticisms of FAccT scholarship to date?''
    \item ``How do you believe the FAccT conference can address the above issues and limitations in the near future?''
    \item ``Please briefly describe how you believe ``impact'' should be defined for FAccT scholarship.''
\end{itemize}}
\noindent Participants also had the option of sharing additional thoughts about FAccT, {further information about} their backgrounds and identities, and feedback about the questionnaire. Additional details about survey design can be found in Appendix \ref{app:sec:survey-validation}, and the full survey is included in Appendix \ref{app:sec:questionnaire}.

\begin{figure*}[b]
\hspace{0.22in}
    \centering
    \includegraphics[width=0.9\textwidth]{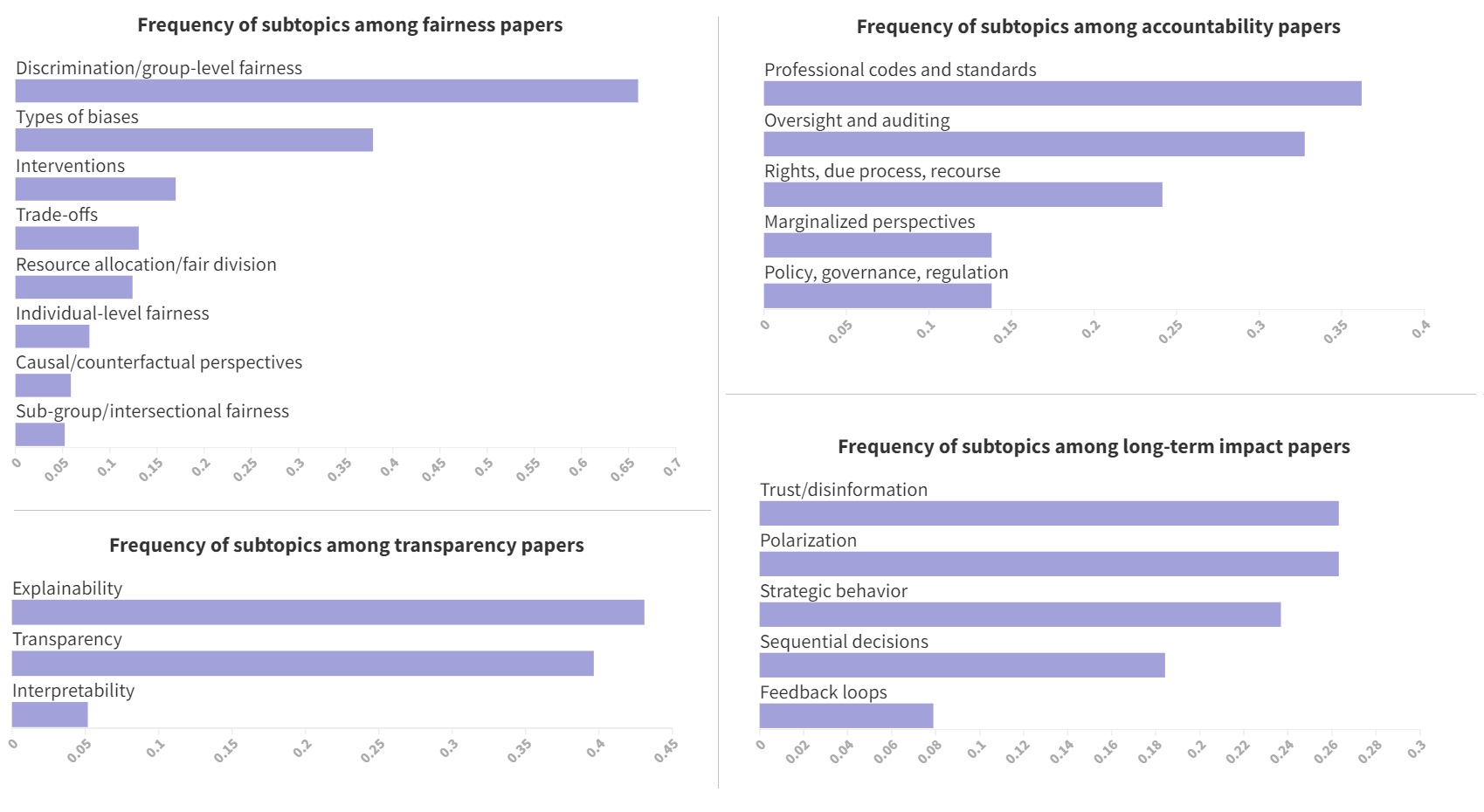}
    \vspace{-0.175in}
    \caption{Relative frequency of subtopics among FAccT papers within 4 topics: fairness, accountability, transparency, and long-term impact.}
    \label{fig:subtopics}
    \vspace{-0.175in}
\end{figure*}

\xhdr{Ethical considerations}
Our study was reviewed and approved by Carnegie Mellon University's {Institutional Review Board (IRB)} prior to its launch. Beyond the standard criteria required to pass institutional review (e.g., obtaining informed consent), we accounted for several additional considerations, including our own {motives} and our participants' goals and aspirations. 
First, our motivation for involving FAccT affiliates in our research was to reflect their voices and opinions about the role and future of FAccT. 
Second, throughout the process of designing and administering the questionnaire, we were keenly aware of our participants' care and investment in the FAccT community and scholarship. As discussed by \citet{howard2019ways}, this awareness heightened our sense of duty to ensure their meaningful, active, and sustained participation in the research process. Toward these goals, we attempted to make the questionnaire more collaborative by asking participants about their {preferred} definitions of ``impact'' and ``value''---two of the key concepts we hoped to evaluate through our study. Additionally, we provided several opportunities for free-form expression of general thoughts and opinions regarding FAccT. Participants also were given the option of continuing their interactions with us (e.g., by emailing the Principal Investigator of the study directly, filling out the feedback textbox on the questionnaire, or expressing interest in participating in one-on-one interviews). Third, we made the questionnaire anonymous by default to prevent biasing participant responses (e.g., in anticipation of their responses being read and interpreted by researchers and later shared with the broader community). Nonetheless, we provided the option of identifying themselves and/or their responses if {a participant so} wished, so that we could name and acknowledge their contributions to our research. 
Fourth, to be mindful of our participants' time, we made all questions optional, but we mentioned that we appreciated their input on as many of them as they believed they were qualified to answer. 
Finally, we weighed the possibility of compensating our participants; considering the nature of their contributions, we concluded that monetary compensation could be perceived as disdainful commodification on our part {and bias the sample}.

\xhdr{Population, sampling, and respondents' demographics} We compiled a list of candidate participants by combining {publicly available data of} FAccT main-track authors, reviewers, and organizing committee members for 2018--2021.\footnote{This list did not represent all conference attendees, authors of rejected papers, and other contributors who were not in the conference proceedings.} 
Out of the 918 individuals emailed, 60 self-selecting FAccT affiliates responded (6.5\% response rate). 
Questionnaire respondents were asked questions about their affiliation with FAccT and demographic information. Among all participants, 44 (75\%) identified with STEM expertise and 23 (39\%) identified with humanities, social sciences and arts (HSA) expertise (some marked both). At least 36 have participated in the FAccT conference as attendees, 39 {as authors, 12 as organizing committee members, and 36 as reviewers}. 54 respondents characterized their political views, of which 67\% marked Liberal, 0 marked conservative or libertarian, and all remaining political views were individually submitted, of which the most-common was `socialist.' 23 (41.8\%) of respondents said that they belong to a marginalized/disadvantaged group, and 32 (58.2\%) stated they did not.

\section{FAccT Research Themes and Topics}\label{sec:themes}

This section describes our thematic investigation of FAccT scholarship. We utilized one qualitative method (manual coding) as well as two quantitative methods (topic modeling and citation network analysis) to extract themes and patterns in FAccT publications.

\xhdr{Data collection}
Our data consisted of text documents--i.e., articles that have been peer-reviewed and published by the FAccT conference in 2018--2021. 
Data collection was straightforward: We downloaded the full conference proceedings through the ACM website on September 25, 2021, with free access through Carnegie Mellon and Cornell University's credentials. In all, we downloaded 224 papers, 186 of which were full-length proceedings articles (the rest are non-archival extended abstracts).  

\subsection{Coding to Identify Topics, Methods, and Applications}
Our qualitative coding aimed at organizing papers to respond to the following questions: What are FAccT's main research \textbf{topics and subtopics}? How have research efforts been distributed among them? 
Which research \textbf{approaches and methods} are frequently utilized? 
Which \textbf{application domains and datasets} have been studied extensively?


\looseness-1\xhdr{Coding process}
We used a combination of \emph{predetermined} and \emph{emergent} codes~\citep{stuckey2015second}. Based on our research questions, our predetermined codes were classified into ``research design'', ``topic'', ``application'', and ``data set''. Within each category, we predetermined several codes as follows: Referring to standard classifications of quantitative and qualitative research approaches~\citep{creswell2017research}, we added 8 qualitative and 6 quantitative research designs as codes under the category of ``research design'' (see Appendix \ref{app:sec:codebook}). 
For ``topics'', based on our initial review of the data, we started with three high-level topics: \texttt{fairness}, \texttt{accountability}, and \texttt{transparency}. 
We used prior FAccT CFPs to determine subtopics under each of these broad topics.
We went through 20\% of papers in our dataset to validate our initial protocol. Accordingly, we added a fourth topic, \texttt{long-term impact}, as an emergent code. Our complete code book and additional details about our coding process are provided in Appendix \ref{app:sec:codebook}.

\begin{figure*}[t!]
\begin{center}
 \vspace{-0.1in}
        \begin{minipage}[c]{.7\textwidth}
        \includegraphics[width=5.45in]{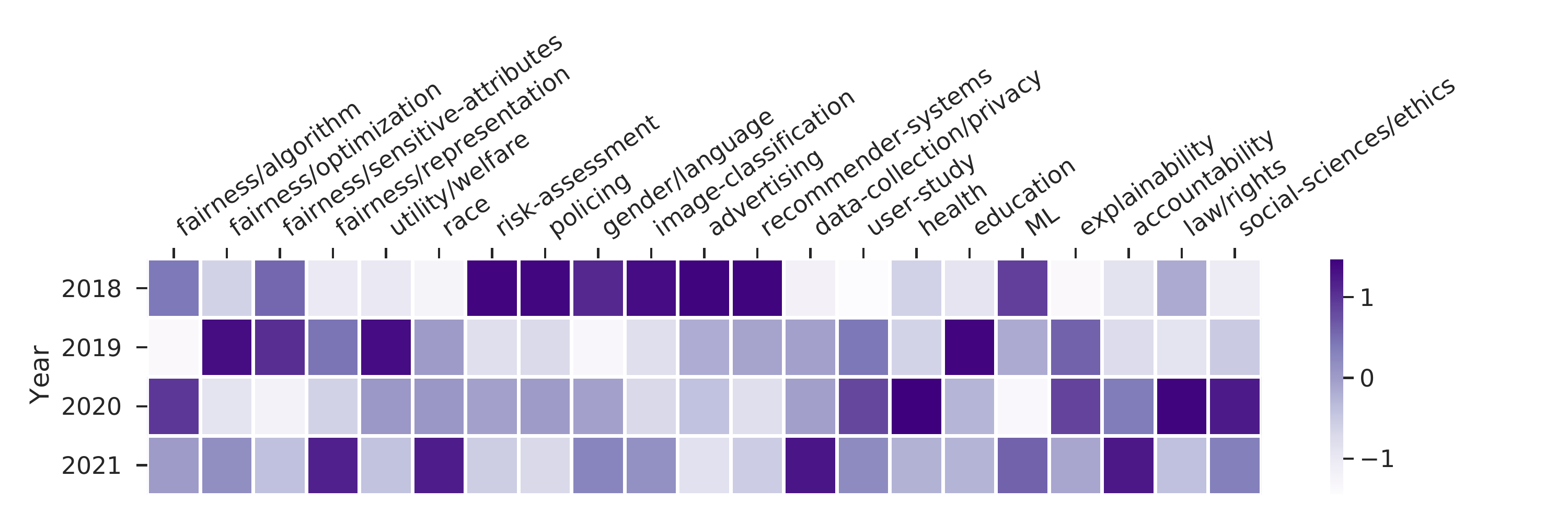}
        
        \includegraphics[width=5.45in]{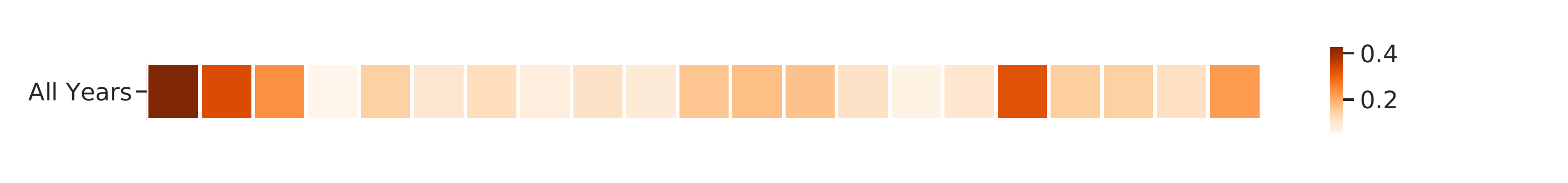}
        \end{minipage}
        \hfill
        \begin{minipage}[c]{.255\textwidth}
        \caption{Topic distributions: normalized, grouped by year (top); across all years (bottom). By year, we see variations over time; e.g., \texttt{accountability} becomes more prevalent. For all years, we see the relative prevalence of topics in general. \texttt{fairness}-related topics and \texttt{ML} dominate all other topics.}
        \label{fig:lda}
        \end{minipage}
\end{center}
\vspace{-0.2in}
\end{figure*}

\xhdr{Findings} Our coding analysis revealed that in terms of the four broad research \emph{topics} identified above, \texttt{fairness} has received the highest level of attention (69\% of all publications), followed by \texttt{transparency} (26\%) and \texttt{accountability} (26\%), and \texttt{long-term impact} (17\%). Other topics (e.g., \texttt{privacy} and \texttt{human factors}) were addressed in 31\% of papers. {Each paper can be categorized under multiple codes, so the percentages do not necessarily add up to 100.} See Figure \ref{fig:subtopics} for a sub-topic break-down of FAccT publications. \texttt{Discrimination/group-level}, \texttt{explainability}, \texttt{professional codes and standards}, and \texttt{trust/disinformation} were the most prevalent subtopics under fairness, transparency, accountability, and long-term impact, respectively.   

In terms of \emph{research design}, 18.9\% of the papers used quantitative empirical methods (e.g., randomized experiments; causal methods) and 32.9\% used qualitative empirical methods (e.g., interviews). 61.3\% were broadly labeled as \texttt{STEM} papers and 29.7\% were labeled as Humanities, Social sciences, and Arts (\texttt{HSA}). Further, we coded 11.3\% of publications under \texttt{philosophy}, 11.3\% as \texttt{professional}, and 9.9\% as \texttt{law}. The top three \emph{datasets} utilized in FAccT publications were \texttt{Adult Income}~\citep{kohavi1996uci}, \texttt{COMPAS}~\citep{larson2016how}, and \texttt{German Credit}~\citep{hofmann1994german}, all publicly available. Of the 75 papers identified as using an `off-the-shelf' dataset, 18.7\% used \texttt{Adult Income} 17.3\% used \texttt{COMPAS}, and 10.7\% used \texttt{German Credit}. A total of 23 papers (10.3\%) used original, empirical datasets (excluding synthetic datasets).

\subsection{Unsupervised Methods to Identify Themes}
We employed two unsupervised approaches to discover themes across FAccT: 1) LDA-based topic modeling~\cite{blei2003lda} on archival FAccT papers, and 2) community detection on a citation network consisting of FAccT publications and their immediate citation connections. In this section, we discuss the two approaches and draw insights from the two models. We provide additional details in the Appendix, and the accompanying code can be found in our \textcolor{blue}{\href{https://github.com/pasta41/facct_retrospective}{online repository}}.

\xhdr{Topic modeling} 
We use Latent Dirichlet Allocation (LDA)~\citep{blei2003lda} to elicit topics from FAccT proceedings. LDA enables us to model each paper as a distribution over (latent) topics, where each topic represents a distribution over vocabulary words. Higher-valued topic weights indicate that a topic is more prevalent in the learned model (Appendix~\ref{app:sec:lda}). We provide two heatmaps to visualize the learned topic distributions in Figure~\ref{fig:lda}: topic weights grouped by FAccT conference year (top), in which we normalize each topic's weights to clarify changes year-over-year (i.e., we subtract a topic's mean over the 4 years and scale by the standard deviation), and topic weights for FAccT overall (bottom). We trained our model using $k=22$ topics and initially examined the unnormalized results year-over-year. These results indicated that that one of the $22$ topics contained words commonly used in the sciences to indicate uncertainty, which dominated over the other $21$ topics; these were stop-word-like words like ``may'', ``should'', and ``possibly.'' For clarity of presentation, we remove this topic from Figure~\ref{fig:lda}. 

        
       
\begin{figure*}[t!]
\vspace{-0.05in}
        \begin{minipage}[c]{.475\textwidth}
        \centering
        \noindent
        \includegraphics[width=3.4in]{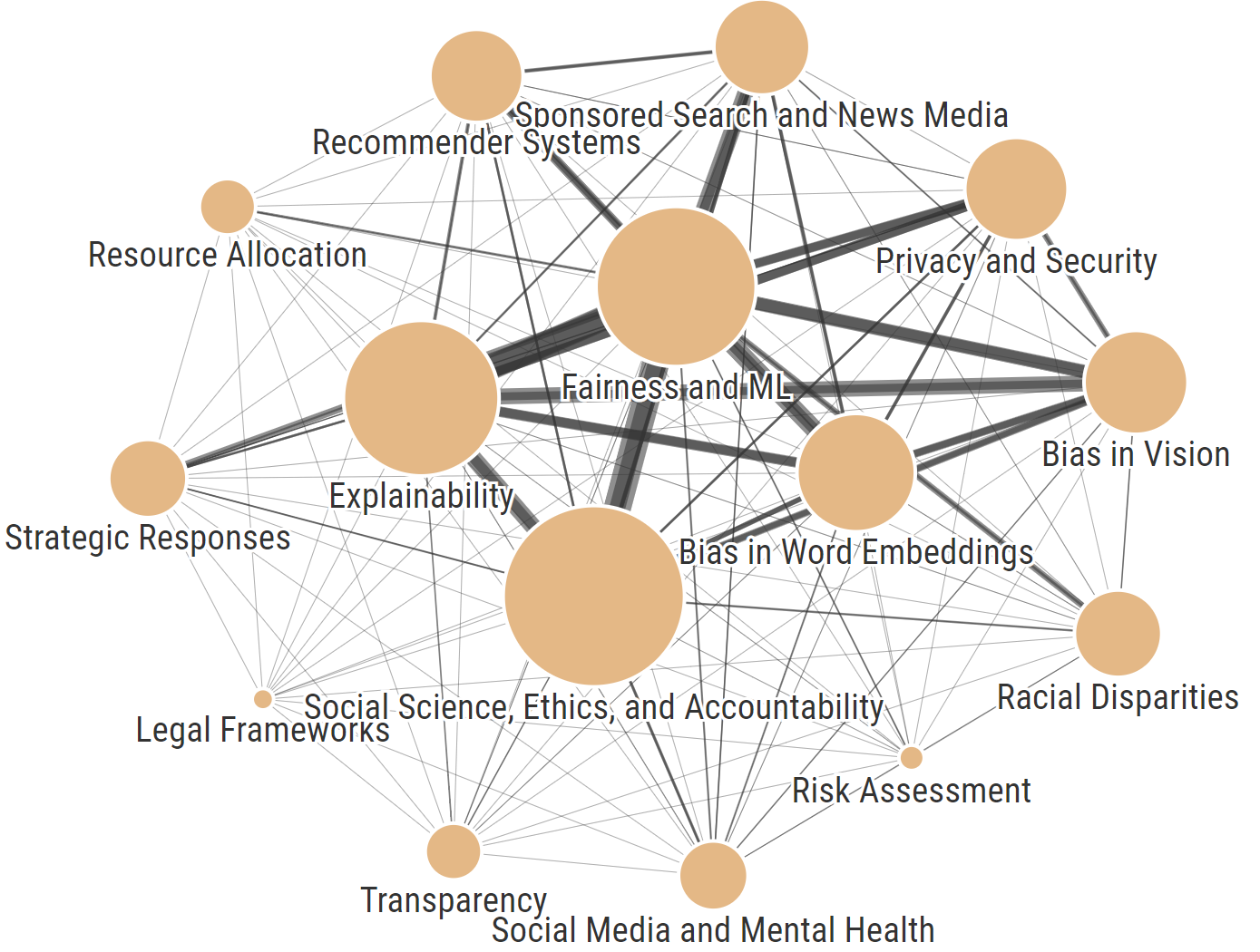}
        \vspace{0.01in}
        \caption{Communities detected using the S2ORC citation network. The size of a given node is representative of the number of papers in the corresponding community. Edge thicknesses represents the volume of citations between the respective communities.}
        \label{fig:community_s2orc}
        \end{minipage}
\hfill
\begin{minipage}[c]{.475\textwidth}
\vspace{-0.01in}
            \centering
                {\small
        \begin{tabular}{lll}
            \toprule \textbf{Community Label} & \textbf{Top Terms and Papers}\\ \midrule
            
            \texttt{Social Sc., Ethics,}  & 
            ethics social accountability \\ \texttt{Accountability} & \citet{selbst2019fairness} 
            \\\hline
            
            \multirow{2}{*}{\texttt{Fairness and ML}}   & 
            fair learning machine \\& \citet{hardt2016equality} 
            \\ \hline
            
            \multirow{2}{*}{\texttt{Explainability}} 
            & explanation learning model \\& \citet{ribeiro2016why} 
            \\ \hline
            
            \texttt{Bias in Word}  &  
            bias embeddings word \\\texttt{Embeddings} & \citet{bolukbasi2016man} 
            \\ \hline
            
            \multirow{2}{*}{\texttt{Bias in Vision}}  & 
            learning bias recognition \\& \citet{buolamwini2018gender} 
            \\ \hline
            \multirow{2}{*}{\texttt{Privacy and Security}}  & 
            privacy data learning \\& \citet{dwork2006calibrating}
            \\ \hline
            
            \multirow{2}{*}{\texttt{Recommender Systems}}  & 
            recommender collaborative diversity \\& \citet{koren2009matrix} 
            \\ \hline
            
            \texttt{Sponsored Search}  & 
            search bias online \\ \texttt{and News Media} & \citet{kay2015unequal} 
            \\ \hline
            
            \multirow{2}{*}{\texttt{Racial Disparities}}  & 
            race ethnicity disparity \\& \citet{hutchinson201950} 
            \\ \hline
            
            \multirow{2}{*}{\texttt{Strategic Responses}}  &
            strategic signaling social \\& \citet{hardt2015strategic} 
            \\ 
            \bottomrule

        \end{tabular}
        }
        \vspace{0.065in}
        \captionsetup{type=table}
        \caption{The ten largest communities. Top terms are selected from the ten most frequent words in paper titles within each community. The referenced papers are the most-cited articles within each community.}
        \label{tab:communities_main}
        \end{minipage}
\vspace{-0.185in}
\end{figure*}

\xhdr{Community detection} We use a variation of the Louvain community detection algorithm~\cite{blondel2008fast} to elicit \textit{communities} within the citation network of all FAccT papers and their immediate neighbors, that is, papers that directly cite FAccT papers or are directly cited by a FAccT paper. (Recall that communities within a network often share common properties or play similar roles within the structure---in our case, they could help us identify research areas and topics.) 
We utilized two datasets for this analysis: the Semantic Scholar Open Research Corupus (S2ORC)~\cite{lo2020s2orc} (Figure~\ref{fig:community_s2orc}) and the AMiner citation network dataset~\cite{tang2008arnetminer}. Both sources provided incomplete citation data, so to ensure the robustness of our findings to the idiosyncrasies of each data source, we analyzed both (Appendix~\ref{app:sec:network}).
The detected \textit{communities} are visualized in Figure~\ref{fig:community_s2orc}. Note that circles represent communitues, or subgraphs of the citation network with high concentrations of citation relationships. 
We named the communities by analyzing the titles of their papers (Appendix~\ref{app:sec:network}).

\xhdr{Findings and discussion} Our topic model identified four fairness-related topics, which we label \texttt{fair\-ness\-/algo\-rithm}, \texttt{fair\-ness\-/opt\-imization}, \texttt{fair\-ness\-/sensitive-attributes} \& \texttt{fair\-ness\-/rep\-resentation}. Papers categorized under all four of these topics contain words like ``fair'' and ``fairness,'' but are separable into distinct themes by the other words that comprise them. For example, \texttt{fairness/sensitive\--attributes} contains attribute-class related words, including ``parity'', ``protected'', and ``sensitive''; \texttt{fair\-ness/\-opt\-imization} has op\-timization\--related words, such as ``optimal'' and ``cost'' (Appendix~\ref{app:sec:lda}). Notably, our results indicate that, separate from the well-documented fact that there are numerous, competing definitions of fairness~\cite{kleinberg2016inherent, kasy2021fairness, cooper2021emergent, binns2018fairness, hanna2020towards, hellman2020measuring}, there are also clear thematic differences in how FAccT papers examine fairness.

The overall results from our topic model (Figure~\ref{fig:lda}, bottom) indicate that fairness-related topics and machine learning (ML) dominate the FAccT proceedings, both within and across years. This point is further supported by our citation analysis, where one of the largest communities we obtained corresponds to \texttt{Fairness and ML}, and other large communities correspond to bias in ML-related areas such as vision and natural language processing. The fact that we obtain these communities related to  \textit{both} fairness and ML (rather than separate communities pertaining to fairness and to ML) suggests not only that these topics are prevalent in FAccT papers independently, but also that fairness \textit{for} ML is a dominant theme across FAccT.\footnote{For one illustrative example of how FAccT scholarship approaches this theme, see \citet{hardt2016equality}, the top-cited paper in the Fairness and ML community.} We observe this result in spite of the more general way that FAccT defines its aims in terms of exploring ``fairness, accountability, and transparency in socio-technical systems''~\cite{facct2022webpage}, which importantly do not require the involvement of ML.

The focus on ML is also indicated by the \texttt{Explainability} community in the citation network. The \texttt{Explainability} community that we obtain focuses on ML models, and while it is less central to the citation network than the \texttt{Fairness and ML} community (as evidenced by the weaker strength of its connections with the other communities), both are similar in size. The focus on ML of both the fairness and the explainability communities is illustrated in Table~\ref{tab:communities_main}, which shows that terms like ``learning'' and ``models'' are among the most frequently occurring terms in paper titles in those communities. Lastly, we also observe from the normalized topic distribution in Figure \ref{fig:lda} (top) that there has been an increase in legal, social science, and ethics papers over the years; thus, while FAccT has predominantly concerned ML papers about fairness, there is evidence it has expanded in scope over the years---so much so that, due to 2021, the overall-FAccT topic distribution in Figure~\ref{fig:lda} (bottom) marks this as one of the most prevalent themes. The citation network in Figure~\ref{fig:community_s2orc} confirms this result; however, it identifies a slightly different theme via a large catch-all community centered around \texttt{Social Science, Ethics, and Accountability}. In the same vein, the topic analysis also shows that \texttt{accountability}, though less of a focus than \texttt{fairness}, has increased in relevance {each year from 2018-2021}.

\section{Values Underpinning FAccT Scholarship} \label{sec:values}

This section examines the values and principles underlying the existing FAccT scholarship. In particular, we ask how prior work has interpreted or operationalized fairness, transparency, and accountability as values. We also ask what other underlying values are represented in FAccT scholarship, and which ones {potentially} deserve further inquiry.

\xhdr{Interpretations of fairness, accountability and transparency}
Within fairness, accountability and transparency, our coding of FAccT publications suggest that these concepts are multi-faceted and the scholarship has focused on specific aspects of these values (see Figure~\ref{fig:subtopics} for the breakdown of these topics into subtopics). In particular, the majority of articles addressing fairness have focused on manifestations of \emph{outcome} disparities across socially salient \emph{groups}. (This is, for instance, in contrast with the limited number of studies dedicated to \emph{procedural} notions of fairness, or those that interpret it as proportionality or reciprocity). Existing work on accountability is mostly centered around \emph{self-governance}---with the bulk of contributions proposing technical auditing tools or industry standards. Prior work has remained largely silent on \emph{regulation} and mechanisms for enforc{ing compliance}.

\subsection{Values beyond Fairness, Accountability and Transparency}
We asked our survey participants \qt{Are there any moral or social values (sufficiently distinct from Fairness, Accountability, and Transparency) that you believe FAccT scholarship should address in near future?} 36 out of 60 participants responded. 

\xhdr{Proposed values, meta-values, and related suggestions} Using manual coding to organize responses, we identified the following values (listed in alphabetical order) as welcome additions to FAccT's scope:
Agency, 
benevolence,
care,
community service,
dignity and respect,
diversity and inclusion,
freedom vs. oppression, 
harm prevention,
intellectual property,
loyalty, 
ownership, 
participation,
privacy,
reflexivity, 
reproducibility,
safety, 
solidarity, 
sustainability, and
truth vs. misinformation. Additionally, participants mentioned the need for broader conceptions of {meta-values}, such as equity, justice, and
trustworthiness. For example, they brought up the necessity of:
\begin{itemize}
    \item Scrutinizing structural facets of justice (e.g., the role of power and capitalism). 
    \item Making room for non-Western values (e.g., indigenous values) and politics. 
    \item Providing mechanisms for striking the right balance between conflicting values and interests (e.g. individuality vs. collectivity, corporate vs. government interests). 
    \item Moving beyond principles and values toward practical implementation.
    \item Providing effective ethical education to AI experts.
\end{itemize}


\xhdr{Moral Foundations Theory (MFT)} 
As an exploratory quantitative approach to understand the moral underpinnings of FAccT and identify commonly-held moral values currently not represented at FAccT, we report the results of an analysis in which we viewed FAccT publications through the lens of the \emph{Moral Foundations Theory (MFT)}\footnote{{We emphasize that our usage of the theory is one among many possible approaches to explore the moral underpinnings of FAccT, and it should not be interpreted as us promoting it over other approaches to mapping the moral landscape (see, e.g., \citep{schwartz1990toward,shweder1997big,fiske1991structures}).}}~\citep{graham2013moral}---a social psychological theory that describes the common bases of people's moral reasoning and tastes as considerations around {several ``foundations,''} including \emph{care}, \emph{fairness and reciprocity}, \emph{loyalty}, \emph{respect for authority}, and \emph{sanctity and divinity}.
{Our analysis utilizes an existing computational tool associated with the MFT:} the Moral Foundations Dictionary (MFD)~\citep{graham2009liberals,frimer2019moral}. The MFD captures the core concepts and terms corresponding to each of the above five foundations. 
The frequencies of MFD terms appearing in the FAccT publications corpus along with the most commonly occurring terms under each foundation can be found in Table \ref{table:most_common}. 

    \begin{table}[tb]
    \caption{The relative frequency of MFD terms in each foundation (out of 48,473 found terms) and the top-5 most frequent terms for each category in FAccT 2018-2021.}
    \label{table:most_common}
    \small
	\begin{tabular}{llp{9pc}}
            \toprule \textbf{Foundation} & \textbf{Freq.}        & \textbf{Most Frequent Terms}  \\ \midrule
            \texttt{care.virtue} & 12.0\% & health protect help share care \\
            \texttt{care.vice} & 4.7\% & harm attack violent vulnerable\newline victim \\
            \texttt{fairness.virtue} & 35.0\% & fair trust equal law justice \\
            \texttt{fairness.vice} & 18.5\% & bias discrimination disparity\newline inequality unfair \\
            \texttt{loyalty.virtue} & 19.1\% & group community company\newline organization  united\\
            \texttt{loyalty.vice} & 0.04\% & outsider betrayed enemy heresy\newline disloyal\\
            \texttt{authority.virtue} & 18.2\% & order protect respect rank police \\
            \texttt{authority.vice} & 0.6\% & illegal  nonconformity unlawful\newline refuse orders \\
            \texttt{sanctity.virtue} & 2.4\% & body organic religion clean faith\\
            \texttt{sanctity.vice} & 1.4\% & drug disease sexual mar pandemic \\
            \bottomrule
            \end{tabular}
                
        \end{table}

\xhdr{Findings and discussion}
Table \ref{table:most_common} provides suggestive evidence on the tendency of FAccT publications to be centered around the first two foundations of MFT, namely, \emph{care} and \emph{fairness}. Prior work has suggested that these two foundations are often emphasized in \emph{liberal} cultures~\citep{graham2009liberals}. Note that while at a first glance foundations such as \emph{loyalty} and \emph{authority} may appear highly frequent in the FAccT corpus, a closer look at the words responsible for those numbers (i.e., `group' and `community' under \emph{loyalty}, and `order' and `protected' under \emph{authority}) suggests that their use in the FAccT context is likely in service of discussing fairness-related concerns. This observation should be contrasted with the concepts such as `patriotism' and `self-sacrifice' for the group or `deference to legitimate authority' and `respect for traditions' that underlie the association of these terms to \emph{loyalty} and \emph{authority}, respectively, in MFD. 
With that in mind, analyzing the terms within the MFD and MFT points to several additional values that are currently not a topic of inquiry at FAccT, including `unity,' `dignity,' `spirituality,' `divinity,' `respect,' `self-determination,' and `freedom.'

\section{The Intellectual and Broader Impact of FAccT Scholarship}\label{sec:impact}

This section aims to characterize the \emph{impact} of FAccT scholarship, in aggregate and across FAccT's topics and domains. To do this, we draw from the participants' responses to our survey. First, we code open-ended responses to our survey to characterize the FAccT community's notion of ``impact'' (Section \ref{subsection:impact-defining}). Second, we use {Likert-scale responses to quantitative questions} to understand views on the following impact-related topics: interdisciplinary exposure and peer review (Section \ref{subsection:interdisciplinarity}), the impact and priority of various research topics (Section \ref{subsection:impact-topic}), and the impact of different application domains (Section \ref{subsection:impact-application}). Finally, we discuss critiques put forward by respondents concerning impact (Section \ref{subsection:impact-criticisms}).

\begin{table*}[b]
\vspace{0.1in}
\caption{Coding and categorization of FAccT affiliates' notions of impact.}
\vspace{-0.02in}
\label{table:impact}
{\small
\begin{tabular}{p{0.2\linewidth} p{0.72\linewidth}}
\toprule \textbf{Code/Category}  & \textbf{Quotes}  \\ \midrule
\raggedright \texttt{Enhancing public literacy and awareness}                     & 
    \qt{Improving awareness of [fairness, accountability, and transparency]-issues in the broader public; improving data \& [AI] literacy of the broader public}, 
    \qt{[Shaping] the broader discourse around what gets built, why, and how.}
             \\ \midrule
\raggedright \texttt{Educating AI experts}                                        & \qt{Ethical frameworks and education of those in the AI community}     \\ \midrule
\raggedright \texttt{Policy influence}                                            & \qt{Real-world influence on [...] policies, [...] software development and data handling standards.}                        \\ \midrule
\raggedright \texttt{Concrete benefits for impacted communities}                  & \qt{Benefit to society from consideration of historically marginalized perspectives/experiences. Applicability of findings/frameworks/techniques to improving technology as deployed in the real world.}              \\ \midrule
\raggedright \texttt{Translation to practice}                                     & Example: Contributing \qt{open source scientific software}                \\ \midrule
\raggedright \texttt{Buy-in from impacted communities}                            & \qt{Enthusiastic participation and research support from communities and identities most likely to experience algorithmic harm under current practices.}                        \\\midrule
\raggedright \texttt{Intellectual paradigm shifts}                                & \qt{[Changing] currently accepted conceptions, terminology, and frameworks}, \qt{Reframing how key stakeholders (decision-makers, policy creators, advocates) understand/think about the world.}                        \\\midrule

\raggedright \texttt{Academic reach}                                             & \qt{Ability to reach different subcommunities in FAccT (CS, Econ, Sociology, etc)} \\\midrule
\raggedright \texttt{Political ramifications}                                     & \qt{FAccT scholarship should not be misused by economic or political stakeholders to calm down upcoming discussions about AI ethics.},  \qt{A particularly negative impact is [the] uptake of technical approaches by industry in ways that amount to empty and detrimental solutionism.} \\
\bottomrule
\end{tabular}
}
\vspace{0.1in}
\end{table*}

\subsection{Defining Impact} \label{subsection:impact-defining}
{This section reports the participants'} views on the ``intellectual merit'' and/or ``broader impact'' of FAccT scholarship across a variety of topics and domains. {For concreteness and grounding, we mentioned} the standardized definitions of these terms proposed by the US government's National Science Foundation (NSF),
which broadly defines \emph{intellectual merit} as the contribution to advancement of knowledge and understanding, and \emph{broader impacts} as benefits to society and contributions to the achievement of specific, desired societal outcomes. (Further criteria and examples can be found in Appendix \ref{app:sec:nsf})
{We also provided participants with the opportunity to define} impact. 
In particular, we asked them to \textit{``briefly describe how [they] believe `impact' \emph{should} be defined for FAccT scholarship.''} {Many} of our participants thought the {NSF} definitions are sufficiently broad. 
Others considered impact to be subjective and difficult to measure for the scholarship as a whole. The rest shared their suggestions on how impact should be defined. Utilizing manual coding, we categorized all responses. The resulting codes and relevant quotes can be found in Table \ref{table:impact}.

\subsection{Analysis of Interdisciplinarity} \label{subsection:interdisciplinarity}

FAccT is a venue which features scholarship from a variety of {disciplines and backgrounds}. The cross-disciplinary topics and research questions that surface in FAccT scholarship require meaningful and constructive communication across these diverse perspectives. We leverage responses to two questions in our survey to characterize how successfully FAccT has facilitated {such} communications.

\xhdr{Findings} In response to our question, \qt{To what extent has the FAccT conference exposed its members to insights and ideas from other disciplines?}, {the mean Likert score across participants was} 3.95 ($\textit{SE}= 0.190, n=58$){, suggesting the general belief that FAccT has positively impacted interdisciplinary interactions.} 
%

{However, one potential challenge for FAccT's interdisciplinary community is its peer review process.}
 We asked participants, \qt{Reviewers of the FAccT conference come from various disciplines and backgrounds. In your experience, how has this impacted the quality of the conference's peer review?}. {Compared to the first question about exposure, respondents' assessments were less positive, with a mean} score of 3.11 ($\textit{SE}= 0.229, n=55$). Notably, these answers did not seem to vary {with participants' background or expertise}.

\xhdr{Community suggestions} {Participants offered several additional}
criticisms and suggestions {concerning the interdisciplinary nature of FAccT}. We classified their {responses} into three categories: 1) {those pertaining to interdisciplinary} communications 2) collaborations and 3) contributions. On the topic of communication, responses pointed out the need for  \textbf{integrating non-STEM perspectives}. For example, one response pointed out a \qt{lack of integration between FAccT and philosophical community (including ethicists, political philosophers), due to the technical nature of FAccT scholarship.}  On the topic of collaboration, one suggestion was to create an \textbf{award for collaborative scholarship}. Other participants expressed \textbf{concerns about quality and rigor} {of contributions that are considered interdisciplinary}. One response stated, \qt{It is not uncommon for individuals to be exposed to a handful of papers, persons, and ideas that are not in their area of expertise, and then think they can produce new scholarship on that basis, because the field encourages `interdisciplinarity.'}
Suggestions related to this criticism emphasize improving the peer review and other organizational aspects of the conference. {For instance, one common suggestion was} separating tracks and reviewing pools based on scholarly expertise. We discuss these ideas in further detail in Section \ref{subsection:venue}.

\begin{figure*}[t!]
    \centering
    \includegraphics[width=\linewidth]{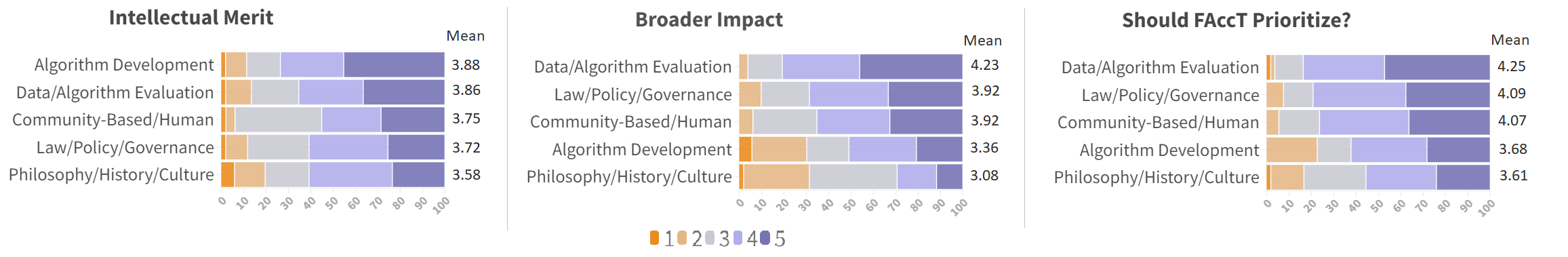}
    \vspace{-0.2in}
    \caption{Participants' views on FAccT scholarship across five topics, specifically their intellectual merit (left), broader impact (middle), and the topics' priority for future FAccT scholarship (right). We illustrate the distribution of respondents' ratings on the 1-5 scale, with 1 corresponding to `very low' and 5 corresponding to `very high.' The percentage of participants who gave a particular rating is represented on the x-axis and mean ratings are reported on the right end of the graphs.}
    \label{fig:topics}
\end{figure*}

\vspace{-2mm}
\subsection{Topic-level Analysis of Impact} \label{subsection:impact-topic}
Survey participants reported their assessment of {intellectual merit, broader impact, and priority of} {FAccT scholarship pertaining to} five {main} topics. The topics were drawn from FAccT's `tracks' listed on the conference CFPs. {Further details about how we chose these topics can be found in Appendix~\ref{app:sec:tracks}. }
We draw a distinction here between \textit{topics} and \textit{application domains}, which we turn to in Section \ref{subsection:impact-application}. In particular, topics are disciplinary, methods-based, or abstract, and they may pertain to any number of application domains.
{Topic-related responses are visualized in Figure \ref{fig:topics}.}

\xhdr{Findings} We highlight {several thought-provoking patterns in participants' responses.} 
First, whereas algorithm development is perceived as a topic {with relatively high} intellectual merit, its broader impact and priority scores are significantly lower, lagging behind three other topics. Second, the philosophy/history/culture topic {may appear} as producing scholarship with comparatively lower merit, impact, and priority. However, {we note that} one potential explanation for this low score {can be perceptions of better/more specialized venues for publishing philosophical contributions outside FAccT. (A similar sentiment came up around topics, such as privacy, during our think-aloud protocols). Related to the importance of philosophical contributions for FAccT, one participant} pointed to problems of \textit{ontology} affecting the quality and rigor of FAccT work. 

\vspace{-2mm}
\subsection{Analysis of Domain-specific Impact} \label{subsection:impact-application}
FAccT scholarship concerns several high-stakes application domains, ranging from criminal justice to education. 
\begin{figure}
    \vspace{-0.1in}
    \begin{center}
        \includegraphics[width = 0.45\textwidth]{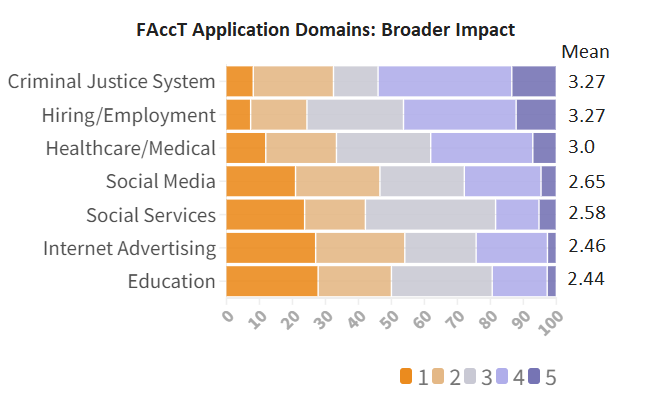}
    \end{center}
    \vspace{-0.1in}
    \caption{Application domains in order of the perceived broader impact of relevant FAccT scholarship (score of 1 corresponds to `very low' and 5 to `very high'). The percentage of respondents that gave a certain rating is represented on the x-axis and mean ratings are reported on the right. Criminal justice system and hiring/employment received the highest scores on average, whereas education and internet advertising received the lowest scores.}
    \label{fig:domains}
    \vspace{-0.1in}
\end{figure}
Focusing on seven key domains which are prominent in FAccT, we asked respondents to characterize FAccT's broader impact. 
(We did not ask respondents to surmise the intellectual merit and priority of these domains, because most of them are known to be of high importance and have significant scholarly attention). Results are reported in Figure \ref{fig:domains}. 

\xhdr{Findings} We see significant imbalances between respondents' ratings of one cluster of domains (criminal justice system, hiring/employment, and healthcare/medial) and the rest (social media, social services, internet advertising, and education). We also find that, even though education was deemed important enough to warrant its own CFP track in 2020, respondents found it to have relatively lower broader impact in comparison to the other domains. 
\vspace{-2mm}
\subsection{Criticisms and Suggestions} \label{subsection:impact-criticisms}
 The responses to survey questions reporting in Figures \ref{fig:topics} and \ref{fig:domains} suggest that there is a rift between the convictions held by community members and the broader impact of some of the work. To explain this rift and find strategies to overcome it, we draw from participants’ open-ended answers about major criticisms and suggestions for FAccT. In particular, we identify a number of barriers to broader impact: (1) Insularity, (2) Narrow Inquiry, and (3) Ontology.

\xhdr{Insularity} A number of respondents pointed out a key {shortcoming of FAccT as a failure to} \qt{address actual problems with impacted communities.} One respondent connected this to \textbf{echo-chamber dynamics} within FAccT. 
Multiple respondents pointed out that FAccT papers overly adhered to \textbf{solutionism}, with one stating that FAccT scholarship has \qt{a tendency to not fundamentally question certain technologies or discuss in what contexts they're (in)appropriate [...] instead [it exhibits] a tendency to incrementally improve such technologies.} 
To overcome these issues, respondents suggested \textbf{engagement with domain expertise} and 
\textbf{public and community engagement}. One respondent recommended that FAccT should \qt{emphasize partnering with translational researchers in real disciplines with `street-level bureaucrats’ and practitioners.} Another broad category of solutions concerned \textbf{diversity and inclusion}. This included a call to expand \qt{Diversity in authors, reviewers, and general participants}, by direct outreach and financial assistance.

\xhdr{Narrow inquiry} A number of respondents pointed out that FAccT scholarship has {had limited} impact because its focus is too narrow or granular. One respondent pointed out the \qt{technical/technocratic approaches to fairness [...] offer a very narrow view of more broadly understood conceptions of fairness in law, policy, society.} Further, respondents warn against the phenomenon of \textbf{fair-washing}, where narrow notions of fairness get used to condone existing practices. Other responses pointed out an over-emphasis on \textbf{Machine Learning}, with one stating \qt{I don't think I've ever read a [FAccT] paper that didn't in some way have to do with ML.} Other responses pointed out {the over-representation of} \textbf{Western- and U.S.-centric} values in FAccT, {which restricts the scope of applicability and impact of FAccT contributions}. One respondent stated: \qt{The conversation in FAccT often assumes a strong familiarity with a set of values and modes of discourse which are currently highly visible in the US, but much less so in other countries, especially non-Western countries. I urge the community to consider what can be done to make researchers who are not fluent in this specific type of discourse feel welcome and able to participate in the community.}

\xhdr{Ontology} Another critical factor that may hinder FAccT’s impact, according to participants, has to do with the conference’s definition of \textbf{key {concepts}}. One respondent stated, \qt{FAccT has not yet settled key ontological questions about the field, leading to incoherent use of key terms, such as `AI' and `algorithm.'} Another respondent noted a lack of \textbf{critical evaluations of mathematical assumptions}, stating that FAccT suffers from \qt{an over-focus on technical results without careful considerations for the mathematical assumptions and constraints used}. One emergent suggestion encouraged \emph{conceptual} work, {in particular, drawing on Science and Technology Studies (STS)}. 


\section{FAccT Community Reflections}\label{sec:community}

This Section aims to shed light on the FAccT \emph{community}'s practices, inter-personal norms, and (academic) culture. The analysis provided here is based on the responses participants gave to our open-ended question inquiring about the most important \emph{criticisms} of FAccT to date and \emph{suggestions} for improvements. Responses covered a wide spectrum of opinions, ranging from broadly positive to highly negative, but the majority of responses lied in the middle of this spectrum--bringing up significant challenges and tensions while recommending steps for improvement. 

\xhdr{Coding process}
42 (out of 60) participants responded to the relevant open-ended questions. 
We followed the process of manual coding proposed by \citet{tesch2013qualitative} to extract patterns from the resulting qualitative data. The output of our coding process for all open-ended responses is summarized in Table~\ref{table:defense} in the Appendix. 
At a high-level, our analysis revealed two major categories of worries concerning the FAccT \emph{community}: one regarding the \emph{organization} and role of FAccT as a conference (including its relationship with industry, government, and traditional disciplines) and another regarding the affiliates' attitudes toward each another and the resulting \emph{culture}.

\subsection{FAccT as a Venue} \label{subsection:venue} 
Participants scrutinized several organizational facets of FAccT as a conference. Major themes reflected in their responses included
\textbf{peer review quality}, which was identified as particularly challenging for FAccT due to its interdisciplinary mission and the varied disciplinary expertise of FAccT reviewers. 
Another concern related to the publication process at FAccT was the risk of \textbf{over-curation} of accepted papers and tutorials, as opposed to \qt{accept[ing] all work that is novel, correct and fits the scope of the conference}. 
Participants offered several suggestions to improve peer review quality. One proposal that appeared several times was \textbf{separating tracks and reviewing pools by expertise.}
One respondent cautioned against this idea, though: \qt{FAccT should strive to be a space that transcends specific disciplinary standards and traditions rather than perpetuating them. We should make more effort to have reviewers from more disciplines rather than creating silos where papers are reviewed only by members of their own disciplines.}
Another proposal was to foster a \textbf{more transparent reviewing process} by clarifying standards and quality measures, possibly revealing the expertise of the reviewers assigned to each submission, and devising an educational/on-boarding process for reviewers.

The second major theme regarded the \textbf{influence of industry} on FAccT as potentially troubling. For example, one participant warned that \qt{The FAccT community should be more careful in how researchers from big tech industry are assigned key positions in the conference.}  
Another described the potential harms of unregulated corporate influence as follows: \qt{It is easy for industry to adopt (or co-opt) some aspects of the work in ways which only minimally help those affected by algorithmic systems, and may even make them worse by giving them a false sense of legitimacy.} Corporate interests can additionally bias the scholarly discourse: \qt{Some questions are not asked or are difficult to ask within industrial research. There is a political bias.}  
As a way of moderating the influence, one participant called on FAccT to \qt{implement stricter funding disclosures for submitted manuscripts, limit corporate researchers' involvement with the [organizing committee], diversify the [organizing committee], promote CRAFT, [clarify that] FAccT is not an auditing organization.}
%
%
Some respondents acknowledged the complex nature of FAccT's connections to industry, and hoped that it can \qt{figure out a way to enable industry to meaningfully contribute (there are many good researchers out there, despite the poor dynamics we've read about in the past year).} 



In addition to above concerns, respondents urged the community to reflect on the potential negative impact of FAccT on standard CS conferences (e.g., by \qt{isolat[ing] people who are concerned with values-oriented work from the main body of the technical community}); the environmental harms of FAccT as a large academic gathering; and the risk or uncertainty of affiliation with FAccT for junior researchers. 

\vspace{-2mm}
\subsection{FAccT's Interpersonal Norms and Practices}
Responses concerning the interpersonal norms and culture of FAccT painted a picture of striking \textbf{divide between several major camps}: \emph{STEM-focused}, \emph{social-scientific}, and \emph{activism-oriented} efforts. One participant described the tension as \qt{between those who see incremental progress as progress and those who believe the only meaningful progress will be revolutionary.} 
On one side of the spectrum, respondents criticized \qt{viewing research with a specific ideological prism rather than a scientific prism.} Some sensed outright \textbf{animosity toward STEM}: \qt{I found myself in several situations where people would speak with true disdain towards engineers, as though this was commonplace and normal.} The participant further expressed shock toward this attitude, especially \qt{for a conference about fairness, accountability and transparency.} The \textbf{lack of inclusivity}
was reflected in calls for increased outreach and tolerance. 

Perceptions of rampant \textbf{non-constructive critisism} were also common among participants. For example, one participant said \qt{FAccT scholarship is all about [criticism] now ... Do we want to be a community of people that \emph{does} stuff or a community of people that \emph{complains} about stuff?} 
%
As a remedy, one participant proposed encouraging \textbf{constructive criticism} by dedicating to it a phase after peer review: \qt{once papers are [accepted], there should be another stage which is constructive criticism where people are invited to challenge the assumptions/values/agendas of the work}.

\section{Conclusion} \label{sec:conclusion}
This reflexive study aimed to shed light on the scholarly field forming around fairness, accountability, and transparency in socio-technical systems. We analyzed the FAccT conference---the \emph{themes} present in its publications,
the \emph{values} that underpin the research, the \emph{impact} of the work, and the culture of its \emph{community}.
Our mixed-methods analysis has used both quantitative and qualitative approaches to study how the FAccT community has directed its emphasis over the past several years. 
Our analysis highlights several significant needs and opportunities in the FAccT community, including (1) further intellectual investment in pressing issues of governance and accountability,
(2) expanding the values underpinning the scholarship, (3) strengthening connections to real-world issues, practices, and stakeholders, and (4) building a more inclusive community. In closing, we hope our contribution benefits the FAccT community by facilitating a constructive dialog around the challenges we face as a diverse, interdisciplinary field aiming to address sensitive, high-stakes socio-technical issues that will only grow in magnitude and significance in the years to come.

\begin{acks}
First and foremost, we acknowledge the vital role that survey participants played in our research. We sincerely appreciate their motivation to benefit the FAccT community by partaking in our study, and we are grateful for their time, mental energy, and commitment to the field. We are grateful to Karen Levy for multiple rounds of invaluable feedback at various stages of this project, as well as Helen Nissenbaum and David Robinson for providing crucial advice.
The authors additionally would like to thank the participants of the AI, Policy,
and Practice (AIPP) group at Cornell, and the Digital Life Initiative (DLI) at Cornell Tech for their thoughtful remarks and suggestions. Last but not least, we are indebted to Madiha Z. Choksi, Emanuel Moss, Emily Tseng, Meg Young, Nil-Jana Akpinar, Amanda Coston, Wesley Deng, Kit Rodolfa, and Joshua Williams for helping to workshop and develop the research instruments for parts of this project.

Benjamin Laufer is supported by DLI at Cornell Tech and NSF CNS-1704527. A. Feder Cooper is supported by AIPP at Cornell University, DLI at Cornell Tech, and the John D. and Catherine T. Mac\-Arthur Foundation. This material is based upon work supported in part by NSF IIS2040929, a
Simons Investigator Award, and a grant from the John D. and Catherine T. MacArthur
Foundation. Any opinions, findings, conclusions, or recommendations expressed in this material are those of the authors and do not necessarily reflect the views of the National Science Foundation and other funding agencies.
\end{acks}

\balance
\bibliographystyle{ACM-Reference-Format}
\bibliography{references}

\appendix

\newpage
\section{Analyzing Themes via Topic Modeling}\label{app:sec:lda}

To discover and analyze themes across the corpus of FAccT publications, we perform topic modeling on the FAccT proceedings from 2018-2021. Our modeling complements authors' self-described keywords and CCS concepts; it enables us to take an unsupervised approach to elicit broader themes in FAccT's conferences. In the analysis that follows, we identify prevalent motifs and temporal patterns, some of which have thus-far remained elusive to the community, and which help provide an understanding of FAccT's incipient disciplinary identity. To elicit topics, we perform Latent Dirichlet Allocation (LDA)~\citep{blei2003lda}. LDA enables us to model each paper as distribution over topics, where each topic representing a distribution over vocabulary words. In other words, at a high-level a topic can be viewed as a set of frequently-present, co-located words; LDA lets us find the topics to which a FAccT paper belongs, based on the individual words that the paper contains. More precisely, for a vocabulary $\mathcal{V}$ and a set of topics $\mathcal{T}$, if a particular word $w \in \mathcal{V}$ has a high probability of being in $t \in \mathcal{T}$, a paper that contains $w$ has a higher probability of being about topic $t$.

\xhdr{Data curation} LDA computes a topic model using a bag-of-words representation of text. We therefore needed to prepare the FAccT proceedings, such that individually-isolated document tokens represented semantically-meaningful words. For example, we removed numerical results tables and math syntax, as individually tokenized numbers and symbols are divorced from their context-specific semantic meaning. We further describe our data curation process in the Jupyter notebook in our \textcolor{blue}{\href{https://github.com/pasta41/facct_retrospective}{online repository}}. This process required a significant effort of code-assisted manual data cleaning, after which we manually verified that the cleaned FAccT corpus preserved the integrity of the original papers' text. We chose to include the $186$ full-length archival proceedings papers to train our model, as a large portion of the non-archival abstracts were not papers, but rather tutorials and other in-conference community-based sessions and performances. We opted not to selectively arbitrate which contributions to include and exclude, and instead included only archival submissions.

\xhdr{Training procedure} The $186$ papers in our dataset constitute a small-text corpus. We therefore looked to prior successful small-text topic analyses to inform our training procedure. Following recently published work~\cite{cooper2021decameron}, we use a Python wrapper of the MALLET library to train our model~\cite{antoniak2021mallet, mccallum2002MALLET}. MALLET, unlike the more-popular Python-based \texttt{gensim} library, uses Gibbs sampling~\cite{geman1984gibbs} for the LDA algorithm's underlying sampling method. Gibbs sampling is an exact Markov chain Monte Carlo technique~\cite{brooks2011handbook}, which \citet{cooper2021decameron} notes has better performance for small-text corpora than inexact, variational-inference LDA implementations~\cite{jordan1999vi}. For the documents submitted to LDA in training, we chunk the FAccT papers into contiguous segments of $200$ words, following the intuition concerning hyperparameter tuning for Gibbs-sampling-based LDA described in the well-cited paper by~\citet{griffiths2004lda}.

\xhdr{Hyperparameter selection}

Choosing the number of topics $k$ for the model to learn requires some domain expertise and some degree of human intervention~\cite{chang2009tealeaves, griffiths2004lda, cooper2021decameron}. We began our experiments with $k=20$, and tried larger and smaller $k$, guided by general advice in ~\citet{griffiths2004lda}, and the results reported reflect those that worked the best when confirming the output topics ($k=22$). From performing this process for the $5$ trained models ($k=19,20,21,22,23$), we selected the model with $k=22$ topics for our analysis. A selection of topic labels, with a subset of topic-specific words, can be found in Table~\ref{table:topics}. 

We chose chunk sizes in such a way that ensured most document lengths are similar: documents are processed as a bag-of-words batch, so using a constant size serves to help normalize. We note that it does not help to chunk based on semantically meaningful sections because LDA operates on a bag-of-words representation, so we lose any semantic relationship between words (or higher level structure, like sections). When removing stop-words, we do not include the resulting chunk for training if it contains fewer than $20$ words. In total, this procedure yields a $7149$-document training corpus, composed using a vocabulary of $23764$ words, with an average of $117.2$ unique words per document.

\xhdr{Additional figures and results tables} We provide comprehensive results in our \textcolor{blue}{\href{https://github.com/pasta41/facct_retrospective}{repository}}. These results include unnormalized results grouped by year, the dataframes used to generate the heatmaps in the paper, as well as by-paper topic distributions for each of the $186$ papers in FAccT. We also include the full topic outs (i.e., the words for each topic).

\begin{table}[ht!]
\caption{A selection of results from our LDA-based topic model: A subset of topics, a corresponding sample of 4 words in the topic, and the (normalized) top-weighted paper for the topic.}
\small
\label{table:topics}
\begin{tabular}{l l l}
\toprule \textbf{Topic Label}        & \textbf{Sample Words} & \textbf{Top Paper}  \\ 
\midrule
\shortstack[l]{\texttt{fairness/}\; \\ 
                           \texttt{optimization}\; \\ 
                           \; \\ 
                           \;} &
\shortstack[l]{optimal \\ 
               fair \\ 
               cost \\ 
               constraint} & 
\shortstack[l]{\citet{ron2021bandits}\; \\ 
                           \; \\ 
                           \; \\ 
                           \;} \\
\midrule

\shortstack[l]{\texttt{futility/}\; \\ 
                           \texttt{welfare}\; \\ 
                           \; \\ 
                           \;} &
\shortstack[l]{group \\ 
               social \\ 
               utility \\ 
               welfare} & 
\shortstack[l]{\citet{heidari2021mobility}\; \\ 
                           \; \\ 
                           \; \\ 
                           \;} \\             
\midrule

\shortstack[l]{\texttt{policing}\; \\ 
                           \; \\ 
                           \; \\ 
                           \;} &
\shortstack[l]{crime \\ 
               allocation \\ 
               police \\ 
               policing} &
\shortstack[l]{\citet{akpinar2021policing}\; \\ 
                           \; \\ 
                           \; \\ 
                           \;} \\                   
\midrule

\shortstack[l]{\texttt{law/rights}\; \\ 
                           \; \\ 
                           \; \\ 
                           \;} &
\shortstack[l]{legal \\ 
               law \\ 
               discrimination \\ 
               rights} & 
\shortstack[l]{\citet{wilson2021screening}\; \\ 
                           \; \\ 
                           \; \\ 
                           \;} \\    
\midrule

\shortstack[l]{\texttt{image-}\; \\ 
                           \texttt{classification}\; \\ 
                           \; \\ 
                           \;} &
\shortstack[l]{images \\ 
               datasets \\ 
               vision \\ 
               face} & 
\shortstack[l]{\citet{yang2020imagenet}\; \\ 
                           \; \\ 
                           \; \\ 
                           \;} \\    
\midrule

\shortstack[l]{\texttt{user-study}\; \\ 
                           \; \\ 
                           \; \\ 
                           \;} &
\shortstack[l]{participants \\ 
               human \\ 
               accuracy \\ 
               decision} & 
\shortstack[l]{\citet{lai2019case}\; \\ 
                           \; \\ 
                           \; \\ 
                           \;} \\    
\midrule

\shortstack[l]{\texttt{health}\; \\ 
                           \; \\ 
                           \; \\ 
                           \;} &
\shortstack[l]{privacy \\ 
               health \\ 
               clinical \\ 
               patients} & 
\shortstack[l]{\citet{suriyakumar2021health}\; \\ 
                           \; \\ 
                           \; \\ 
                           \;} \\    
\midrule

\shortstack[l]{\texttt{fairness/}\; \\ 
                           \texttt{sensitive-}\; \\ 
                           \texttt{attributes}\; \\ 
                           \;} &
\shortstack[l]{fairness \\ 
               protected \\ 
               sensitive \\ 
               attribute} & 
\shortstack[l]{\citet{raz2021independence}\; \\ 
                           \; \\ 
                           \; \\ 
                           \;} \\    
\midrule

\shortstack[l]{\texttt{fairness/}\; \\ 
                           \texttt{representation}\; \\ 
                           \; \\ 
                           \;} &
\shortstack[l]{fairness \\ 
               representativeness \\ 
               moral \\ 
               representative} & 
\shortstack[l]{\citet{malgieri2020gdpr}\; \\ 
                           \; \\ 
                           \; \\ 
                           \;} \\    

\bottomrule
\end{tabular}
\end{table}

\section{Citation Network Analysis}\label{app:sec:network}

Here, we describe in detail our data extraction process [\ref{app:sec:net:data}], the community detection algorithm we use [\ref{app:sec:community_detection}], and the method we use to identify themes from communities [\ref{app:sec:net:themes}].

\subsection{Citation Data}\label{app:sec:net:data}
For our citation analysis, we are interested in observing all articles appearing in FAccT proceedings and all papers that cite FAccT papers or are cited by FAccT papers. We use 2 datasets: the AMiner citation network dataset \citep{tang2008arnetminer} and the Semantic Scholar Open Research Corupus (S2ORC) \citep{lo2020s2orc}, to verify the robustness of our method. Each dataset has unique drawbacks: the AMiner data is incomplete in its resolution of citation links and does not give us complete lists of references to/from FAccT papers, while S2ORC had its latest release in 2020 and does not include FAccT 2021 papers. For the papers from 2018 to 2020, however, S2ORC resolves references better than AMiner.

\subsubsection{AMiner Dataset:} We use the 13th iteration of the AMiner dataset (released May 2021). Each entry in the dataset includes article metadata and contains fields for author list, venue, year of publication, and referenced articles. We obtain the subset of papers relevant for our analysis in three steps. We use the venue field of the dataset to filter in the papers published in FAccT. Since some papers have missing venue fields, as a second step, we manually search for missing papers using string search on words from paper titles or author names across the entire AMiner dataset. The first two steps give us a total of 207 FAccT papers. For the third step, we iterate over the entire dataset and extract articles that either cite, or are cited by, the articles in our seed set. 

\subsubsection{Semantic Scholar Open Research Corpus (S2ORC) Dataset:} S2ORC is similar in metadata format to AMiner and we perform a similar 3-step procedure as described in the section on AMiner. Since S2ORC was released in 2020, we extract data and citation links corresponding only to the first three editions (2018-2020) of the conference.

\subsection{Community Detection}\label{app:sec:community_detection}
For both of our datasets, we construct a directed citation network from the extracted set of papers (if an article A cites an article B, this is represented by a directed edge from A to B). We then identify communities within the resulting network. Here, we describe the algorithm we use to detect communities: a variation \citep{dugue2015dirlou} of the Louvain algorithm \citep{blondel2008fast} that incorporates edge direction in its optimization objective.

The Louvain algorithm and its variations rely on the idea of Modularity ($Q$), which is defined as the fraction of edges that fall within communities minus the expected fraction of such edges if they were distributed at random. More formally, modularity is given by the following expression:

\begin{equation}\label{modularity}
    Q = \frac{1}{2m} \sum_{ij} \left[A_{ij} - \frac{k_ik_j}{2m} \right] \delta_{c_i, c_j}
\end{equation}

Here, A is the adjacency matrix of the graph ($A_{ij}$ is 1 if there exists an edge between i and j and 0 otherwise), $m$ is the total number of edges in the network, $k_i$ is the degree of vertex $i$, and $\delta$ is the Kronecker delta function, which takes the value 1 if both its arguments are equal and 0 otherwise. 

An algorithm such as Louvain with modularity $Q$ as its optimization objective disregards edge direction in a directed graph. It is, in fact, possible (and common practice) to disregard edge directions and to instead use the communities obtained from the corresponding undirected graphs, as is done by the Louvain algorithm. However, incorporating direction information can lead to the identification of more coherent communities as described by \citet{leicht2007commdir}, who propose a formulation of \textit{directed} modularity, that can be used with the Louvain algorithm to incorporate edge direction information while detecting communities.

The change in the optimization objective can be motivated as follows: traditional (undirected) modularity values are high when a \textit{statistically surprising} fraction of edges in a network fall within the chosen communities. \citet{leicht2007commdir} extend this idea further to motivate their definition for directed modularity by suggesting that \textit{any} statistically surprising configuration should contribute to an increase in modularity.  Next, they consider two vertices---A and B---in a directed network such that A has a high out-degree and a low in-degree and B has a high in-degree and a low out-degree. In such a case, A$\,\to\,$B edges are more likely than B$\,\to\,$A edges. Therefore, B$\,\to\,$A edges are more statistically surprising, and should contribute more towards modularity. 

Intuitively, in the case of citation networks, $A$ corresponds to an article with a small number of citations while $B$ corresponds to an article with a larger number of citations. In this case, a B$\,\to\,$A edge ($B$ citing $A$) suggests more strongly that both $A$ and $B$ address the same area than an A$\,\to\,$B edge does.

\citet{leicht2007commdir} formalize this by proposing a measure that can be used as an equivalent of modularity in directed graphs. This quantity, which we denote $Q'$ is given by the following:

\begin{equation}
    Q' = \frac{1}{m} \sum_{ij} \left[A_{ij} - \frac{k_i^{in} k_j^{out}}{m} \right] \delta(c_i, c_j)
\end{equation}

Here, in addition to the symbols introduced in Equation \ref{modularity}, $k_i^{in}$ and $k_j^{out}$ are the in-degree and out-degree of vertices. This is the optimization objective we use while applying the Louvain method to perform community detection.

\subsection{Identifying Themes \& Naming Communities} \label{app:sec:net:themes}
To identify the areas or research directions each of the communities deals with, we study the titles of all papers present in a given community. Concretely, we compute unigram frequencies of terms appearing in paper titles and observe the top 10 terms to estimate the themes that the community deals with. We use both raw and weighted unigram-frequencies. We calculate top terms using two methods: 1) using the raw unigram frequency of terms appearing in the titles and 2) weighting term frequencies by the in-degree of the paper they appear in. 

We get very similar results from our analysis on both datasets. We present the additional results from the AMiner dataset in our GitHub \textcolor{blue}{\href{https://github.com/pasta41/facct_retrospective}{repository}}.

\section{NSF Impact Definitions}\label{app:sec:nsf}

In our questionnaire, we adopted the terms `intellectual merit' and `broader impact' from NSF (National Science Foundation), which defines:

\begin{itemize}
    \item \textbf{Intellectual merit} as the contribution to advancement of knowledge and understanding. (Criteria include sound rationale and reasoning motivating the research,  presenting creative, original, or potentially transformative concepts/approaches, well-organized execution of the research, positive scholarly impact within or outside the field). 
    \item \textbf{Broader impacts} as benefits to society and contributions to the achievement of specific, desired societal outcomes. (Examples include empowering disadvantaged or marginalized individuals and communities; improving equity of access to opportunities; improving literacy and engagement of researchers, practitioners and the public).
\end{itemize}

\section{Survey Development}\label{app:sec:survey-validation}

In this section we briefly describe the development and validation process in designing, piloting, and deploying our questionnaire. The process unfolded over numerous stages of exploratory work, group feedback sessions, in-person interviews and think-aloud protocols, followed by our final iteration of view solicitation: a survey sent out to all FAccT authors with publicly available emails.

We began by drafting a set of relevant research questions and survey questions, which we presented to our research groups in two sessions. These sessions helped identify additional research inquiries, methods, sources of confusion, and relevant similar studies. After updating our questionnaire, we did two rounds of pilot surveys using think-aloud protocols \cite{eveland2000examining, ericsson1998study, whitney1996think, fonteyn1993description}, where the two researchers each did (approximately) half-hour-long think-aloud procedures while subjects filled out the questionnaire, followed by half-hour-long interviews soliciting participants' views on the questionnaire. After conducting these four interviews, the questionnaire was updated to reduce ambiguities in how respondents interpret questions and control questionnaire length. The second round of pilots consisted of three hour-long interviews of the same nature, which were significantly more consistent and led to only minor language changes in the questionnaire.

Once finalized, the questionnaire was sent to every author of a FAccT conference publication, whose email was available either on the FAccT paper or on a personal website. In addition to attaching the questionnaire, we invited participants to participate in extended interviews. After receiving only three volunteers, however, we decided to forego interviews. We made this decision in order to protect the privacy of the research subjects and because we would not have been able to reach saturation with only three interviews.

 \section{Choosing 5 FAccT Topics from Tracks}\label{app:sec:tracks}
 
 We provide more granular details about how the five ``topics'' in our questionnaire (as seen in Figure \ref{fig:topics}) were chosen from the history of tracks in FAccT CFPs. We aggregated these tracks over all years, therefore, the topics chosen do not directly or exhaustively represent these tracks, but rather are winnowed from the full set. We note that the topics do not exactly match the ``tracks'' for two reasons: First, FAccT's tracks varied in terminology and subject year-over-year. Second, the pilot survey procedure, which consisted of 7 think-aloud interviews,  uncovered certain tracks that led to inconsistency and confusion. In particular, the `Security and Privacy' track was omitted from this analysis, because it significantly derailed survey responses during pilots. These participants became confused because they weren't aware of a security and privacy track.

\newpage\section{Coding of FAccT Publications}\label{app:sec:codebook}

Below we include a table representing the exhaustive set of codes used to analyze four years of FAccT papers. Two authors (Laufer and Heidari) were responsible for coding and used the papers' full text. Codes were not mutually exclusive; papers were assigned all relevant codes. 

\begin{table*}[b]
\caption{Codes used to categorize and analyze FAccT papers.}
\label{table:codebook}
\normalsize
\begin{tabular}{l l l}
\toprule \textbf{Category}        & \textbf{Code Type} & \textbf{Code}  \\ \midrule
Title & Open-ended                    & e.g., ``A Statistical Test for Probabilistic Fairness'' \\ 
\midrule
Synopsis & Open-ended & e.g., ``Statistical hypothesis test for unfair classifiers''  \\ 
\midrule
Qualitative Research Design & Categorical & Scholarship review and critique    \\
 &  &  Phenomenology    \\
 &  &  Ethnography    \\
 &  &  Case Study    \\
 &  &  Grounded Theory    \\
 &  &  Narrative    \\
 &  &  Historical    \\
 &  &  Action    \\
 &  &  Discourse Analysis    \\
\midrule
Quantitative Research Design & Categorical & Experimental    \\
 &  &  Survey    \\
 &  &  Causal-comparative    \\
 &  &  Correlational    \\
 &  &  Cross-sectional    \\
 &  &  Longitudinal    \\
\midrule
Field & Categorical & STEM    \\
 &  &  Law    \\
 &  &  Philosophy    \\
 &  &  Social sciences and humanities    \\
\midrule
STEM Contributions & Categorical & Algorithm development    \\
 &  &  Evaluations, metrics, measures    \\
 &  &  Mathematical models and analysis    \\
 &  &  Package, library, toolbox    \\
\midrule
Topic: Fairness & Categorical & Discrimination/group-level (un)fairness    \\
 &  &  Individual-level (un)fairness    \\
 &  &  Subgroup/intersectional (un)fairness    \\
 &  &  Causal/counterfactual perspectives    \\
 &  &  Tradeoffs    \\
 &  &  Interventions and algorithms    \\
 &  &  Types of biases    \\
 &  &  Resource allocation/fair division    \\
\midrule
Topic: Transparency & Categorical & Transparency (audit, reproduce, data sharing, proprietary)    \\
 &  &  Explainability (human-understandable translation)   \\
 &  &  Interpretability (as inherent feature of model/algorithm)    \\
\bottomrule
\end{tabular}
\end{table*}

\onecolumn
\begin{table}[ht!]
\caption{Codes used to categorize and analyze FAccT papers - Continued.}
\label{table:codebook2}
\normalsize
\begin{tabular}{l l l}
\toprule \textbf{Category}        & \textbf{Code Type} & \textbf{Codes}  \\
\midrule
Topic: Accountability & Categorical & Human rights and freedoms, due process, recourse    \\
 &  &  Policymaking, governence, and regulatory frameworks   \\
 &  &  Professional codes, institutional procedures, industry standards    \\
 &  &  Oversight and auditing mechanisms, compliance, liability    \\
 &  &  Models from historically marginalized perspectives    \\
\midrule
Topic: Long-term/social impact & Categorical & Strategic behavior and its consequences    \\
 &  &  Sequential decisions and interventions    \\
 &  &  Feedback loops    \\
 &  &  Polarization    \\
 &  &  Trust/disinformation    \\
\midrule
Topic: Others & Categorical & Privacy, profiling, surveillance    \\
 &  &  Human factors   \\
 &  &  Other desiderata    \\
\midrule
Applications/Domains & Categorical & Internet Advertising    \\
 &  &  Recommendation Systems   \\
 &  &  E-commerce    \\
 &  &  Social Media    \\
 &  &  Entertainment and Media    \\
 &  &  (Criminal) Justice System    \\
 &  &  Law enforcement and policing    \\
 &  &  Lending    \\
 &  &  Healthcare/medical    \\
 &  &  Hiring/employment    \\
 &  &  Social services    \\
 &  &  Design and Robotics    \\
 &  &  Computer Vision Software    \\
\midrule
Off-the-shelf datasets & Categorical & COMPAS    \\
 &  &  Crime and Communities   \\
 &  &  Adult income    \\
 &  &  German credit   \\
 &  &  FICO    \\
 &  &  HMDA    \\
 &  &  MovieLens    \\
 &  &  IMDB    \\
 &  &  LSAT    \\
 &  &  Student    \\
 &  &  MNIST    \\
 &  &  CIFAR-10    \\
\midrule
Original Datasets & Open-ended & e.g., ``US EPA risk assessments''  \\ 
\midrule
Type of AI & Categorical & NLP  \\ 
 &  &  Vision    \\
 &  &  Deep Learning    \\
 &  &  Systems (PL, DL, ...)    \\
\midrule
Misc Notes/Questions & Open-ended &   \\ 

\bottomrule
\end{tabular}
\end{table}

\newpage 
\onecolumn
\section{Coding of Open-ended Survey Responses} \label{sec:app:suggestions}

\begin{table}[H]
\caption{Coding and categorization of FAccT affiliates' opinions, criticisms, and suggestions for improvement.}
\label{table:defense}
\footnotesize
\begin{center}
\begin{tabular}{l l l  lll}
\toprule
\textbf{Applicability} &   \textbf{Category}   &   \textbf{Criticisms/Codes}  &   \textbf{Suggestions}    \\
\midrule
\multirow{1}[1]{*}{Conference}   &   \multirow{1}[1]{*}{Organization\:\:} &
            \shortstack[l]{Peer review quality\; \\ 
                           \; \\ 
                           \; \\ 
                           \; \\
                           \; \\
                           \; \\
                           \; \\
                           \; \\
                           \;}    &  
            \shortstack[l]{Separate track for HSA, STEM, and non-academic work \\ 
                           Separate reviewing pools \\ 
                           Reward truly interdisciplinary contributions \\ 
                           Add a stage after acceptance for constructive criticisms \\ 
                           Clarify standards for (interdisciplinary) contribution \\ 
                           Make it more transparent \\ 
                           Design education/onboarding process for reviewers}    \\
    \cmidrule{3-5}
    & & Over-curation &   Move toward more inclusive acceptance criteria \\
    \cmidrule{3-5}
    & & \shortstack[l]{Industry influence\; \\ 
                           \; \\ 
                           \; }    &  
            \shortstack[l]{Clarify conflict-of-interest policy for authors \\ 
                           Implement funding disclosures \\ 
                           Limit corporate involvement in conf. organization}    \\
    \cmidrule{3-5}
    & & \shortstack[l]{Misc.\; \\ 
                           \;}    &  
            \shortstack[l]{Don't change CFP and deadlines \\ 
                           Consider carbon footprint of conference}    \\
\midrule
\multirow{3}[4]{*}{Conference}   &   \multirow{3}[4]{*}{Relations\:\:} &
            Industry    &  &    \\
    \cmidrule{3-3}
    & & Public sector and governance    &  &  \\
    \cmidrule{3-3}
    & & Other disciplinary venues    &  & \\
        
\midrule
\multirow{6}[4]{*}{Community}   &   \multirow{6}[4]{*}{Culture\:\:} &
            \shortstack[l]{Activism-oriented vs. scientific agendas\; \\ 
                           \; \\ 
                           \;}    &  
            \shortstack[l]{Acknowledge the value of non-technical approaches \\ 
                           Acknowledge the value of math models/solutions \\ 
                           Clarify the academic, audit, \& activist role of FAccT}    \\
    \cmidrule{3-5}
    & & Lack of inclusivity &   Promote CRAFT \\
    \cmidrule{3-5}
    & & Non-constructive criticism &   & \\
    \cmidrule{3-3}
    & & Ingroup-outgroup dynamics &   & \\
    \cmidrule{3-3}
    & & Animosity toward STEM &   & \\
    \cmidrule{3-3}
    & & Misc. (risk for junior researchers) &   & \\

\midrule
\multirow{3}[4]{*}{Community}   &   \multirow{3}[4]{*}{Interdisciplinarity\:\:} &
            \shortstack[l]{Collaborations\;}    &  
            \shortstack[l]{Emphasize translational research}    \\
    \cmidrule{3-5}
    & & \shortstack[l]{Communications\;}    &  
        \shortstack[l]{Develop shared conceptual infrastructure/vocabulary}    \\
    \cmidrule{3-5}
    & & \shortstack[l]{Contribution and novelty\;}    &  
        \shortstack[l]{Clarify standards for (interdisciplinary) contribution}    \\

\midrule
\multirow{6}[4]{*}{Scholarship}   &   \multirow{6}[4]{*}{Insularity\:\:} &
    Lack of practical impact &   Engage w/ ``street-level bureaucrats'' \& practitioners \\
    \cmidrule{3-5}
    & & Lack of diversity and inclusion &   Outreach to communities (e.g., financial assistance) \\
    \cmidrule{3-5}
    & & Intellectualism / echo-chamber dynamics &   & \\
    \cmidrule{3-3}
    & & Solutionism &   & \\
    \cmidrule{3-3}
    & & Lack of engagement with domain expertise &   & \\
    \cmidrule{3-3}
    & & Lack of public and community engagement &   & \\

\midrule
\multirow{2}[2]{*}{Scholarship}   &   \multirow{2}[2]{*}{Narrow inquiry\:\:} &
            \shortstack[l]{Fairness\; \\ 
                           \; \\
                           \; \\
                           \;}    &  
            \shortstack[l]{Broaden inquiry to notions of justice \\ 
                           Address systemic oppression \\ 
                           Include more STS contributions}    \\
    \cmidrule{3-5}
    & & Machine learning &   & \\
    \cmidrule{3-3}
    & & Western and US-centric values &   & \\
    \cmidrule{3-3}
    & & Risk of ``fair-washing'' &   & \\
    
\midrule
\multirow{3}[4]{*}{Scholarship}   &   \multirow{3}[4]{*}{Quality \& rigor\:\:} &
    Ontology &   Encourage conceptual work \\
    \cmidrule{3-5}
    & & Lack of critical eval. of math assumptions &   & \\
    \cmidrule{3-3}
    & & Discouraging mathematical contributions &  & \\
\bottomrule
\end{tabular}
\end{center}
\end{table}

\newpage\section{Questionnaire Copy}\label{app:sec:questionnaire}
\centering
\fbox{\includegraphics[scale=0.43]{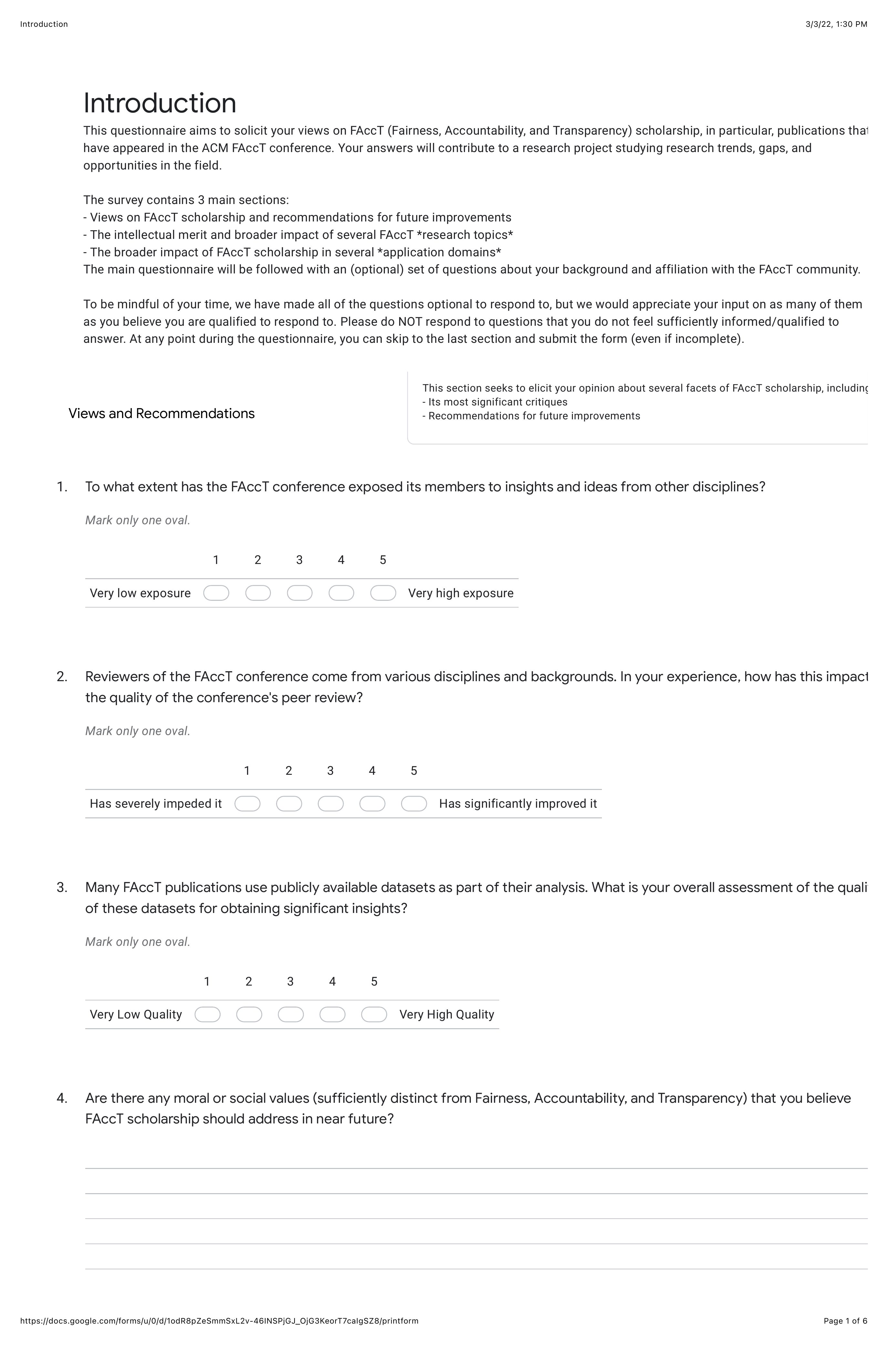}}\newpage

\fbox{\includegraphics[scale=0.45, page = 2]{figure/FAccT_perspectives_appendix.pdf}}\newpage

\fbox{\includegraphics[scale=0.45, page = 3]{figure/FAccT_perspectives_appendix.pdf}}\newpage

\fbox{\includegraphics[scale=0.45, page = 4]{figure/FAccT_perspectives_appendix.pdf}}\newpage

\fbox{\includegraphics[scale=0.45, page = 5]{figure/FAccT_perspectives_appendix.pdf}}\newpage

\section{Consent Form for Survey Respondents}
\centering

\fbox{\includegraphics[scale=0.675]{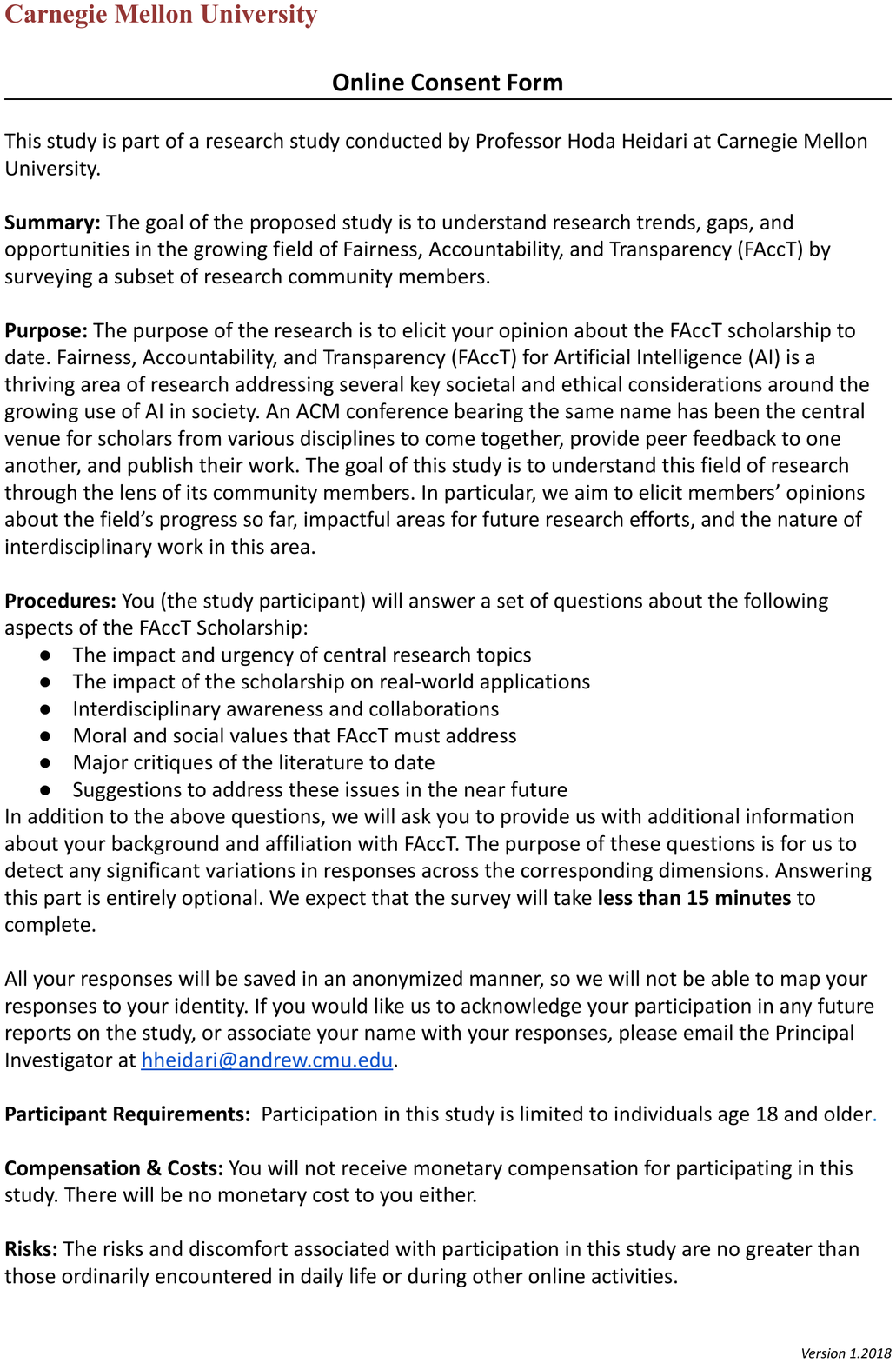}}\newpage

\fbox{\includegraphics[scale=0.675, page = 2]{figure/Online-consent.pdf}}

\end{document}